\documentclass[11pt]{article}
\usepackage[utf8]{inputenc}
\usepackage{amsmath}
\usepackage{amsfonts}
\usepackage{amssymb}
\usepackage{graphicx}
\usepackage{verbatim}
\usepackage{bm}
\usepackage{url}
\usepackage{algorithm2e}
\usepackage{color}
\usepackage{lscape}

\newtheorem{theo}{Theorem}[section]

\newtheorem{definition}[theo]{Definition}

\begin{document}

\title{Liquidity commonality does not imply liquidity resilience commonality: A functional characterisation for ultra-high frequency cross-sectional LOB data}
\author{Efstathios Panayi\thanks{Department of Computer Science, WC1E 6BT, London, UK. Corresponding author. Email: efstathios.panayi.10@ucl.uk. } \; Gareth Peters\thanks{UCL, Department of Statistics, WC1E 7HB, London,  UK} \; Ioannis Kosmidis \thanks{UCL, Department of Statistics, WC1E 7HB, London,  UK}\\
}
\maketitle

\begin{abstract}
We present a large-scale study of commonality in liquidity and resilience across assets in an ultra high-frequency (millisecond-timestamped) Limit Order Book (LOB) dataset from a pan-European electronic equity trading facility. We first show that extant work in quantifying liquidity commonality through the degree of explanatory power of the dominant modes of variation of liquidity (extracted through Principal Component Analysis) fails to account for heavy tailed features in the data, thus producing potentially misleading results. We employ Independent Component Analysis, which both decorrelates the liquidity measures in the asset cross-section, but also reduces higher-order statistical dependencies. 

To measure commonality in liquidity resilience, we utilise a novel characterisation proposed by \cite{panayi2014market} for the time required for return to a threshold liquidity level. This reflects a dimension of liquidity that is not captured by the majority of liquidity measures and has important ramifications for understanding supply and demand pressures for market makers in electronic exchanges, as well as regulators and HFTs. When the metric is mapped out across a range of thresholds, it produces the daily Liquidity Resilience Profile (LRP) for a given asset. This daily summary of liquidity resilience behaviour from the vast LOB dataset is then amenable to a functional data representation. This enables the comparison of liquidity resilience in the asset cross-section via functional linear sub-space decompositions and functional regression. The functional regression results presented here suggest that market factors for liquidity resilience (as extracted through functional principal components analysis) can explain between 10 and 40\% of the variation in liquidity resilience at low liquidity thresholds, but are less explanatory at more extreme levels, where individual asset factors take effect.
\end{abstract}


\section{Introduction}
\label{sec:intro}
A modern challenge in financial econometrics is to summarise and study statistical features, or characteristics, of large-scale datasets, derived from unevenly-spaced observations at an ultra high-frequency rate. The massive data structure outlining the buying and selling interest in an asset is known as the Limit Order Book (LOB), and the challenge is to assess such data structures for a large number of assets, over an extended period time. In this regard, we present a novel perspective on understanding aspects of liquidity in the equity asset space and its cross-section, for a large pan-European electronic equity exchange. 

The fluctuation of market liquidity has been suggested to originate from the adverse selection problem faced by market makers more than four decades ago \cite{bagehot1971only}. Since then, a rich literature has developed in single-asset liquidity, examining both the properties of liquidity measures in and of themselves, but also the effects of liquidity on asset pricing. As examples of the latter, we mention \cite{amihud1986asset}, who model the effects of the spread on asset returns and find evidence of a `liquidity premium', with assets with higher spreads commanding higher returns. \cite{amihud2002illiquidity} confirm that a return–illiquidity relationship exists over time, but \cite{constantinides1986capital} finds less of an impact in multi-period models. 

Since these earlier studies, access to massive high-frequency limit order data has allowed for significantly larger studies on liquidity across markets to be undertaken. This is a topical aspect of big data analysis in financial econometrics, and efforts to improve the understanding of liquidity evolution and co-evolution in the asset cross-section have recently been a focal issue of studies in equities, \cite{hasbrouck2001common,karolyi2012understanding,riordan2013public,sklavos2013liquidity}, commodities and futures \cite{frino2014commonality,marshall2013liquidity} and foreign exchange and bond markets  \cite{holden2014empirical}. 

There are several facets to this problem that are of interest, including how to define and quantify for these datasets different aspects of liquidity, as well as how to measure temporal commonality of such liquidity measures in the cross-section of assets. To this end, recent studies have focused on summarising the massive LOB data sets for each asset over time through either parametric time-series models, empirical estimates of liquidity measures or some combination of both, see for instance \cite{marshall2013liquidity}, \cite{frino2014commonality} and \cite{sklavos2013liquidity}. 

Recent studies involving different assets for data from both primary and secondary exchanges have shown that for a number of different liquidity measures, one may observe a high degree of commonality in liquidity or, as we will show in this paper, commonality in illiquidity. In addition, there is documented evidence for some degree of time scale invariance in the commonality of liquidity \cite{righi2014liquidity}. For equities, the observed commonality is significant at both the market and the industry level \cite{chordia2000commonality,huberman2001systematic}, while \cite{brockman2009commonality} also provide evidence of this, in both developed and emerging markets. The degree of commonality found across multiple different exchanges led the latter to comment that `firm-level liquidity cannot be understood in isolation'. Liquidity co-movement has been found to be prevalent particularly during equity market breaks and debt market crises \cite{hasbrouck2001common}.

The impetus for our line of research comes from the aforementioned literature, where commonality is quantified specifically via a projection based analysis, e.g. through principal component methods. \cite{korajczyk2008pricing} provides such a study for the equity market, while \cite{mancini2013liquidity} finds even greater commonality in foreign exchange rate liquidity. Typically, such studies involve first obtaining (or approximating, if detailed LOB data is unavailable) a liquidity measure over a particular period, e.g. the time series of concurrent measurements of the inside spread for a number of assets over a month. From these time series, one then obtains the first few principle components (PCs), which reflect the dominant modes of market wide liquidity behaviour. The time series observations for the liquidity of individual assets are then regressed against the market factors (given by the PCs), over the entire period. High values for the coefficient of determination in the regressions for different assets then indicate commonality in their liquidity. 

Our first contribution is to show that, at least in the equity space, the assumption that one can capture all the features of the liquidity commonality via the sample covariance of liquidity measures for each asset, as in the PCA regression approach, will not always be appropriate. In particular, using only second order moments will not capture heavy tailed features observed in the empirical distribution of the liquidity for certain assets. The outcome of using only PCA methods, which are based on second moments, is that the analysis is then driven by the most illiquid assets, which act as outliers in the cross-sectional dataset. We therefore perform a projection-pursuit based ICA (Independent Component Analysis), which addresses this issue by incorporating higher order information, to assess commonality in liquidity for our equity data for both the spread and XLM liquidity measures.    

In \cite{kyle1985continuous}, the notion of what comprises a liquidity measure is considered and the three core aspects that were identified were consequently adopted in the subsequent literature on this topic. These aspects are tightness, depth and resiliency. The commonality in the first two of these concepts have been considered in the aforementioned analyses on liquidity commonality (by quantifying commonality in the spread or depth of the LOB). The third aspect of liquidity, relating to resilience, is more difficult to address. Recently, in \cite{panayi2014market}, a quantitative measure of resilience for any liquidity measure has been developed, based on standard survival regression models. The authors define the TED (Threshold Exceedance Duration) metric as the duration of liquidity deviations from a particular liquidity threshold. Such an approach then allows one to obtain TED information for different thresholds, and construct a curve of the expected TEDs as a function of the threshold, which they termed the Liquidity Resilience Profile (LRP). 

Since LRPs are informative about the level of LOB liquidity replenishment for each asset, a commonality analysis can identify clusters of assets for which we would expect a swift return to a high liquidity levels after a shock. To the extent that liquidity replenishment in the millisecond environment is predominantly the domain of high-frequency market makers, LRPs can also indicate the presence or absence of such traders in particular assets. This is in line with the theoretical predictions of \cite{foucault2005limit}, who suggested that large spreads would be more common in markets dominated by impatient traders (those submitting aggressive market orders, rather than passive limit orders). Using the LRP curve as our building block, in this paper we extend the literature on liquidity commonality by quantifying commonality in liquidity resilience for two common liquidity measures, namely, the inside spread and the Xetra Liquidity Measure (XLM), a cost-of-round-trip type measure.


The LRP curves for 82 European stocks traded on the Chi-X on a particular day are first obtained through a smoothed functional data representation, in order to reduce the high dimensionality of the liquidity resilience data. The market factors contributing to the variation in daily liquidity resilience may then be obtained through a functional principal component analysis (FPCA) of the LRP curves. We demonstrate that there is a consistency in the shape of the first three functional principal components (FPCs) over time and then regress the LRPs for individual assets against these FPCs, interpreting their explanatory power as a measure of commonality in the liquidity resilience profile of the given asset with the market factors. 

Functional linear regression yields a functional coefficient of determination, for the continuum of the domain of the function. This is a valuable in our analysis, as one can establish the range of liquidity thresholds for which a particular asset's resilience behaviour has commonalities with market factors. For the equities dataset under consideration, we found that the first 3 FPCs could explain between 10 and 40\% of the variation in liquidity resilience at low liquidity thresholds. However, at more extreme liquidity exceedance thresholds, the commonality between individual asset liquidity resilience profile behaviour and market factors diminishes significantly, and individual asset factors take effect. 

Our second contribution in this paper therefore fits into the extant liquidity commonality literature as a special case, as we investigate whether market resilience can explain liquidity replenishment behaviour in individual assets. We demonstrate that to undertake such a study one may now utilise the LRPs for a range of assets each day. This is important, as resilience is inextricably related to the rate of replenishment of orders after a liquidity shock, and as such, it must also be associated with the presence of liquidity providers in a particular market. Identifying commonality in liquidity resilience can therefore help one understand which assets are the focus of high-frequency market making strategies, and under what market conditions.

This paper is structured as follows: Section \ref{sec:liqintro} provides an overview of the extant literature in liquidity and commonality in the equities and FX asset classes. Section \ref{sec:liqcommonality} quantifies the degree to which past findings are reflected in our data. Section \ref{sec:liqres} introduces the notion of liquidity resilience and the proxy studied in this paper. Section \ref{sec:fdaintro} introduces functional data analysis and its inherent advantages for modelling high-frequency, irregularly spaced financial data, and Section \ref{sec:lrpfits} explains how the functional LRP representation is obtained from discrete TED duration data. Section \ref{sec:fpca} delineates how we determine the main modes of variation in the daily LRP curves through functional principal component analysis. Section \ref{sec:fpcaregression} presents the results of the functional principal component regression and interprets the findings. Section \ref{sec:conc} concludes.

\section{Liquidity in high frequency data}
\label{sec:liqintro}

A number of liquidity measures, or proxy substitutes that approximate these measures, have been proposed, and these generally reflect one or more of the following aspects \cite{kyle1985continuous}:
 \begin{enumerate}
 \item Tightness, `the cost of turning around a position over a short period of time'
 \item Depth, `the size of an order flow innovation required to change prices a given amount'
\item Resiliency, 'the speed with which prices recover from a random, uninformative shock.
 \end{enumerate}

Besides the ubiquitous bid-ask spread and variants thereof, discussed in detail by \cite{holden2014empirical}, examples of liquidity/illiquidity measures include price impact \cite{brennan1996market}, price reversal due to temporary price impact (studied in \cite{stambaugh2003liquidity}), and the ratio of average absolute returns to trading volume \cite{amihud2002illiquidity}. It should be noted that in many of these studies, liquidity measures are obtained by approximation, due to the great expense required to obtain highly detailed LOB data. An approximation for the spread, for example, can be obtained via the method of \cite{roll1984simple}, using the first-order serial covariance of price changes. In measuring commonality, it is also common to obtain lower frequency proxies for liquidity (\cite{amihud2002illiquidity,stambaugh2003liquidity}), in order to reduce the data to a manageable size. 

\cite{goyenko2009liquidity} demonstrates the superiority of high-frequency liquidity benchmarks, compared to low-frequency proxies, while \cite{mancini2013liquidity} argues for the use of good quality data as a necessity for measuring the determinants of liquidity. In this paper, we use a millisecond-timestamped dataset to obtain liquidity data over a four month period, and through the reconstruction of the LOB, we do not need to rely on approximations of liquidity measures. This gives us the advantage of being able to obtain very accurate estimates of liquidity, and draw clear conclusions on liquidity commonality and liquidity resilience commonality.

\subsection{Liquidity commonality in the asset cross-section and its implications}
The majority of the literature discusses single asset liquidity and thus only captures individual variation in liquidity dynamics. Recently, however, there has been a burgeoning interest in studying the cross-sectional variation in liquidity in a number of assets over a period of time. One of the earliest studies to consider the co-movement of liquidity was in the work of \cite{chordia2000commonality}. This was achieved through a simple parametric model setting, by regressing liquidity changes for each asset against market or industry liquidity changes. The authors identified asset specific and aggregate market trading levels as being amongst the determinants of individual asset liquidity. Liquidity co-movement was assumed to result from the risk of maintaining inventory in the presence of institutional funds with correlated trading patterns.

Since this study, a number of asset specific and liquidity measure specific studies have been developed to quantify commonality. \cite{hasbrouck2001common} adopted a distinct framework, using a combination of PCA and canonical correlation analysis to study the commonality in liquidity measures for the Dow 30. This uncovered the most important across-asset common factors in the price discovery/liquidity provision process in equity markets. They found that both returns and order flows are characterised by common factors. The liquidity measures they considered included variants of the spread, depth and the ex-ante trading cost. They found that commonality in order flows can account for roughly two thirds of the commonality in returns, but the common factors in the liquidity proxies above are relatively low. 

Adding to the findings of \cite{chordia2000commonality}, the study of \cite{domowitz2005liquidity} also demonstrated that liquidity commonality in the Australian Stock Exchange may be induced by the co-movement in supply and demand, which materialises in the LOB as the cross-sectional correlation in order types (market and limit orders). The economic justification for this order type co-movement stems from traders' efforts to minimise execution costs, by submitting limit orders in an illiquid market, and market orders in a liquid one. They then demonstrated a linkage between liquidity commonality and return co-movement, which they argued is a key component of portfolio selection. Interestingly, they also argued that in contrast to liquidity commonality, return commonality is less affected by the correlation of order types, but is more related to the co-movement of aspects of the order flow, and specifically, order direction and size. 

Due to the recent developments in big data analytics and the increasing availability of data, the processing of massive, multi-asset, multiple-day high frequency LOB datasets has become more tenable. Consequently, there has been an increasing interest in extending the smaller studies discussed above to encompass multiple days, assets and exchanges. Studies such as those by \cite{brockman2009commonality} begin to employ similar techniques to identify possible commonality across massive datasets. In this study, they consider 47 markets (exchanges) in 38 countries. In addition to the exchange-level commonality identified by \cite{chordia2000commonality}, they find a global component in bid-ask spreads and depths, as well as regional components.

Similarly, an analysis of a massive dataset of more than 4000 firms over almost 20 years by \cite{korajczyk2008pricing} found that approximately 50\% of the time-series variation in firm-level quoted and effective spreads can be explained by the first 3 principal components, and their results suggest liquidity risk as being a priced factor in stock returns. More recently, \cite{karolyi2012understanding} considered daily equity data for 21,328 stocks in 40 developed and emerging countries between 1995 and 2004. They are able to demonstrate commonality in returns, liquidity and turnover and they explain this commonality by features on both the supply and demand sides of the market. On the supply side, they considered factors relating to funding liquidity of financial intermediaries and on the demand side, factors related to investor protections, investor sentiment and trading behaviour of institutions.

In the FX space, utilising a dataset considering 40 FX rate liquidities over an extended period of 20 years, \cite{karnaukh2013understanding} found that commonality can explain an average of 36\% of the variation in liquidity.  However, this is higher in currencies in developed countries, as well as in times of market distress. The computational constraints of undertaking analysis across such an extended period of time are handled by extracting individual FX rate liquidity through PCA across the three best low frequency liquidity proxies (the ones most highly correlated to high frequency proxies). The authors find that co-movements of FX rate liquidities are strong for at least the last 20 years, and, certainly, significantly stronger that in the equities asset class. 

Understanding liquidity commonality is crucial for the success of strategies like the carry trade. For 9 currency pairs, \cite{mancini2013liquidity} document strong contemporaneous comovements across exchange rate liquidities, and extract common information across 5 different liquidity measures. In common with the work presented in this paper, they are able to utilise a high quality dataset and thus do not rely on approximating measures of liquidity to perform their analysis. They use both averaging (used previously by \cite{chordia2000commonality} and \cite{stambaugh2003liquidity} and PCA (used by \cite{hasbrouck2001common} and \cite{korajczyk2008pricing}) to extract market-wide liquidity. They test for commonality by regressing individual liquidity measures against the first component for every exchange rate, and find that this explains between 70 and 90\% of the variation. 

\subsection{Notation and liquidity measures}
In general we will reserve upper case letters to denote random variables, bold for random vectors and lower case letters for the realizations of these random variables and vectors. In addition we utilise the following notation for a single asset, on a single trading day.

\begin{itemize}
\item $a$ denotes the ask, $b$ denotes the bid

\item $P_{t}^{b,i}$ $\in\mathbb{N}^{+}$ is a random variable (RV) of the price of the $i^{th}$ level bid at time $t$ 

\item $P_{t}^{a,i}$ $\in\mathbb{N}^{+}$ is an RV of the price of the $i^{th}$ level ask at time $t$

\item $TV_{t}^{b,i}$ $\in\mathbb{N}^{n}$ is an RV for the volume of orders (in terms of the number of shares) at the $i^{th}$ level bid at time $t$  

\item $LM_t$ is an RV at time $t$ for the generic proxy of the liquidity measure. 

\end{itemize}

In this paper we consider the following commonly used measures of liquidity: the \underline{inside spread} given by $LM_t:= S_t=P^{a,1}_{t}-P^{b,1}_{t}$; and the \underline{Xetra Liquidity Measure (XLM)}
{\small{
\begin{equation*}
\begin{split}
LM_t:=XLM_t(R) &= \frac{\sum _{i=1}^{k}TV_{t}^{a,i}(P_{t}^{a,i}-P_t^m)+(R-\sum ^{i=1}_{k}TV_{t}^{a,i})(P_{t}^{a,k+1}-P_t^m)}{R} \\
& \;\; + \frac{\sum _{i=1}^{k}TV_{t}^{b,i}(P_t^m-P_{t}^{b,i}) + (R-\sum ^{i=1}_{k}TV_{t}^{b,i})(P_t^m-P_{t}^{b,k+1})}{R}
\end{split}
\end{equation*}
}}
where we set $R=\min(25000,\sum_{i} TV_{t}^{a,i},\sum_{i} TV_{t}^{b,i})$, i.e. the minimum of 25000 of the local currency (GBP, EUR or CHF) and the volume available on either side of the LOB.

In terms of the aspects of liquidity delineated above, the inside spread reflects the tightness aspect, while the XLM, as a Cost of Round Trip (CRT) measure, is also descriptive of the depth of volume at each level of the LOB. Whereas $R$ is fixed in the definition of the measure on the Xetra exchange\footnote{\url{http://xetra.com/xetra/dispatch/en/xetraLiquids/navigation/xetra/300_trading_clearing/100_trading_platforms/100_xetra/600_xlm}}, we allowed it to vary, so that the measure was still defined when there is insufficient volume in the LOB. 

\section{Liquidity commonality in a secondary market (Chi-X): PCA, ICA and regression}
\label{sec:liqcommonality}
The dataset that we consider here is from Chi-X, a secondary pan-European equity multilateral trading facility (MTF), which operates as a pure LOB. It contains the entire LOB activity for a large number of assets, for the period between January and April 2012. That is, it includes every limit order submission, execution and cancellation in that period for each asset, some of which consist of several hundred thousand events daily, and with timestamps that are accurate to the millisecond. We can then rebuild the LOB from `flat' order files, in order to extract liquidity measures and other variables of interest without resorting to approximations. 

We first revisit the standard PCA regression approach for liquidity commonality analysis for a selection of 82 of the most liquid stocks on Chi-X, from 3 countries (UK, France and Germany) and 10 different industries, for which further information is provided in Appendix \ref{sec:comp}, Table \ref{tab:assets}. In order to perform such an analysis, one has to ensure that the liquidity measurements for all assets are at regular intervals throughout the trading day. Since the LOB for each asset is an event-driven stochastic process, we need to sample the process to obtain evenly-spaced observations of liquidity, in order to perform a PCA analysis. We thus perform a pre-processing step, first constructing the LOB for each asset and then sampling to obtain liquidity measurements aligned at 1 second intervals. We thus obtain measurements of the spread and XLM every second, for all 82 assets, throughout the 4 month period under consideration.  

We then perform PCA on the liquidity data every day, and regress the liquidity of individual assets against the first 3 PCs (which we consider to be the market factors). This is similar to the analysis undertaken, e.g. in \cite{mancini2013liquidity}, and allows us to investigate temporal commonality in liquidity throughout a day. Performing a daily regression (rather than a regression over the entire period as in \cite{mancini2013liquidity}) enables us to assess the fit over time and identify features that would otherwise be lost through time averaging and smoothing the signal. 

\begin{figure}[ht!]
	\begin{center}
			\hspace{-3.5em}
	\includegraphics[height = 4cm, width=0.42\textwidth]{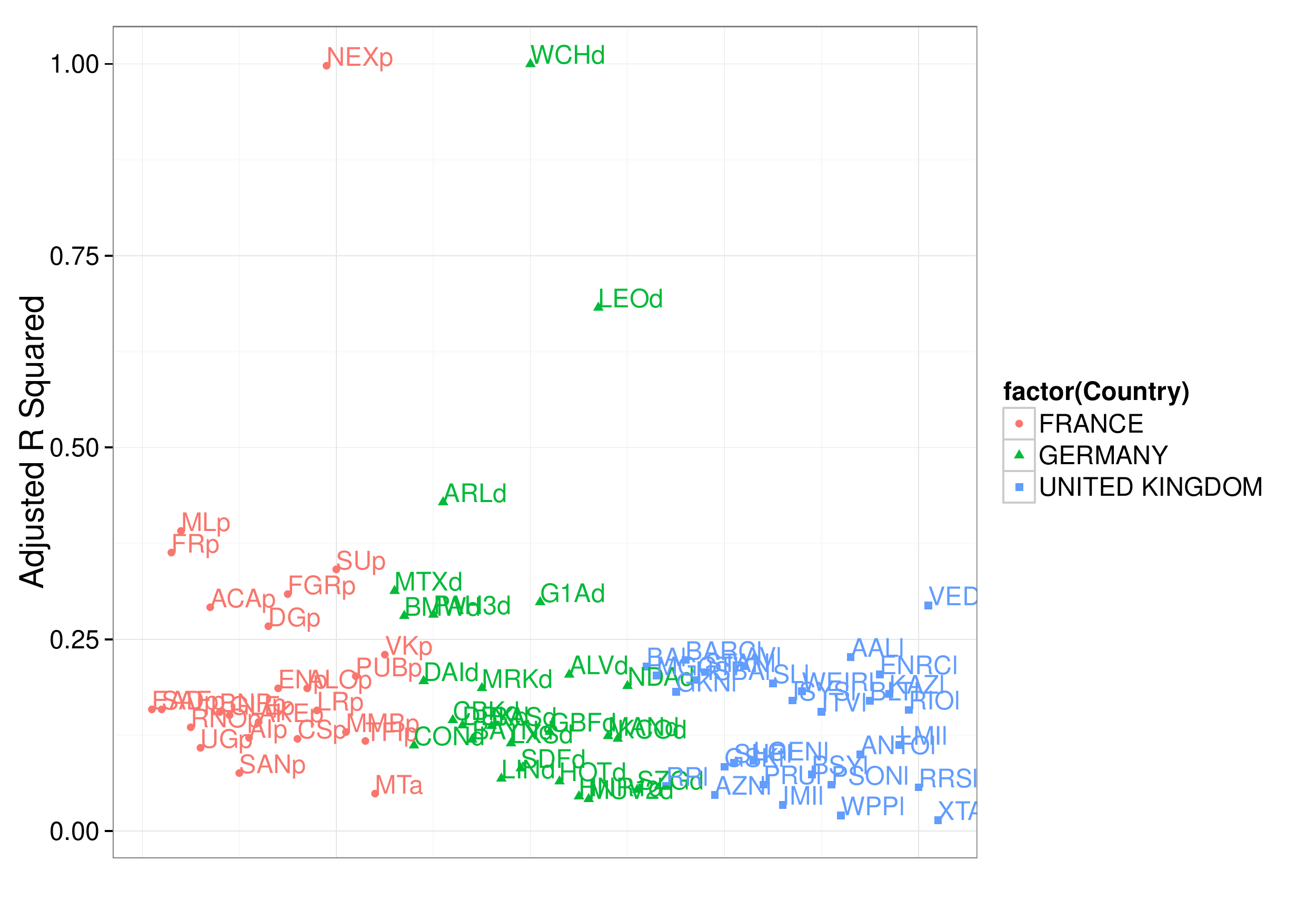}
			\hspace{3em}
		\includegraphics[height = 4cm, width=0.42\textwidth]{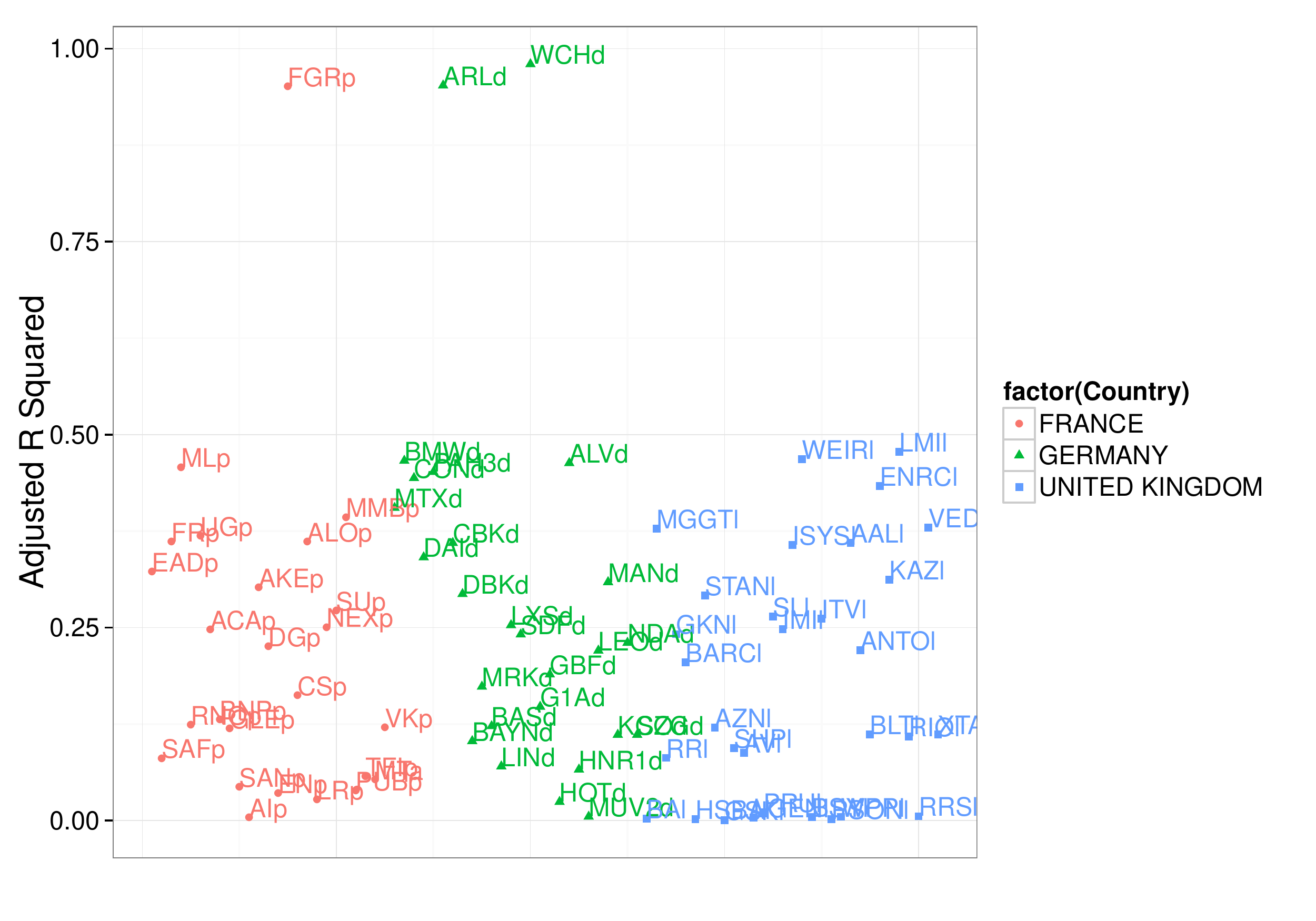}
			\includegraphics[height = 4cm, width=0.49\textwidth]{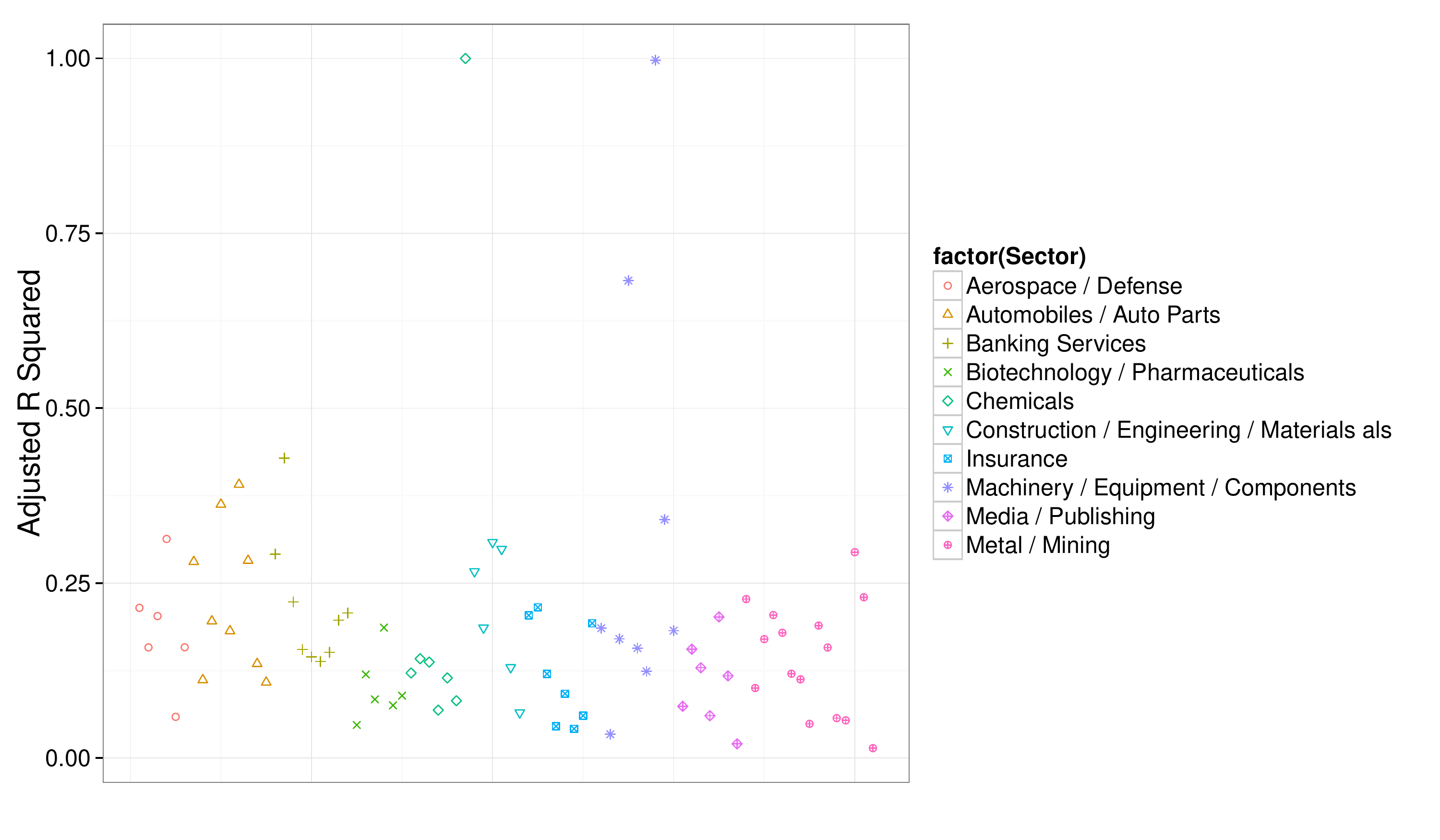}
		\includegraphics[height = 4cm, width=0.49\textwidth]{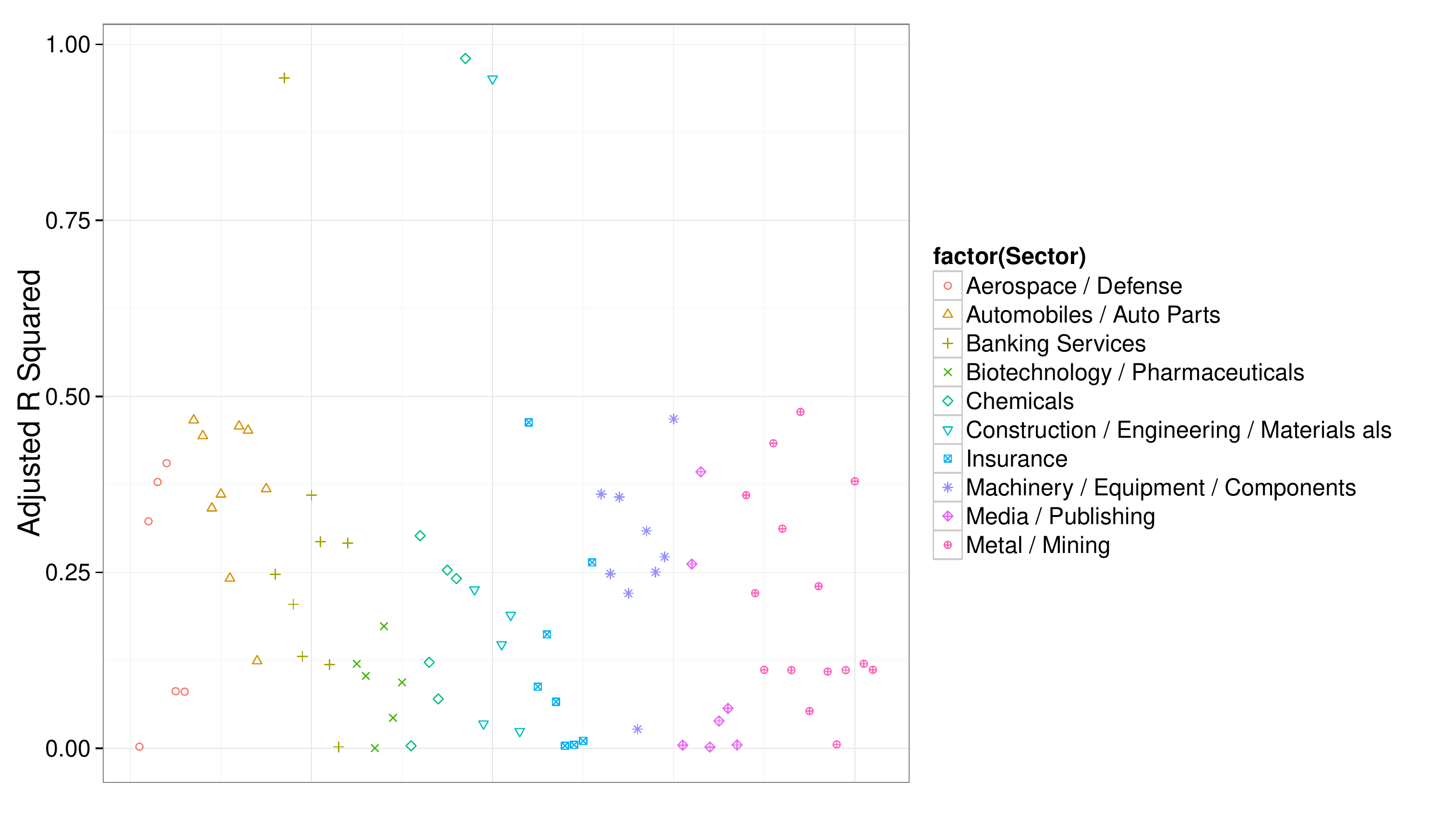}
	\caption{The $R^2$ values obtained from regressing individual asset liquidity against the first three PCs obtained across assets for the spread(left) and XLM(right), where assets are grouped by country(top) and sector(bottom). The labels indicate the Chi-X symbol of every asset.}
	\label{fig:regpca}
	\end{center}
\end{figure}

Figure \ref{fig:regpca} shows the $R^2$ scores of the regression for every asset on a randomly selected day, where the assets are broken down by country and by sector. For most assets and for both the spread and the XLM, the $R^2$ score is around 25\%, although we notice that there are particular assets (e.g. NEXp, WCHd) which have very high $R^2$ scores. 

Similar results are observed consistently throughout the 4 month period. A summary of the explanatory power of the PCA regression for every asset over time is provided in Figure \ref{fig:allr2pca}. The high $R^2$ scores for certain assets imply that the first few `market' PCs for the liquidity measures, which are obtained from the cross-section of all assets on a given day, essentially mirror the liquidity of these few particular assets. That is, the PCA and resulting commonality analysis is driven by those assets. This is a feature we would like to understand, as it has not been discussed previously in the literature. 

\begin{figure}[ht!]
	\begin{center}
	\includegraphics[height = 3.5cm,width=0.4\textwidth]{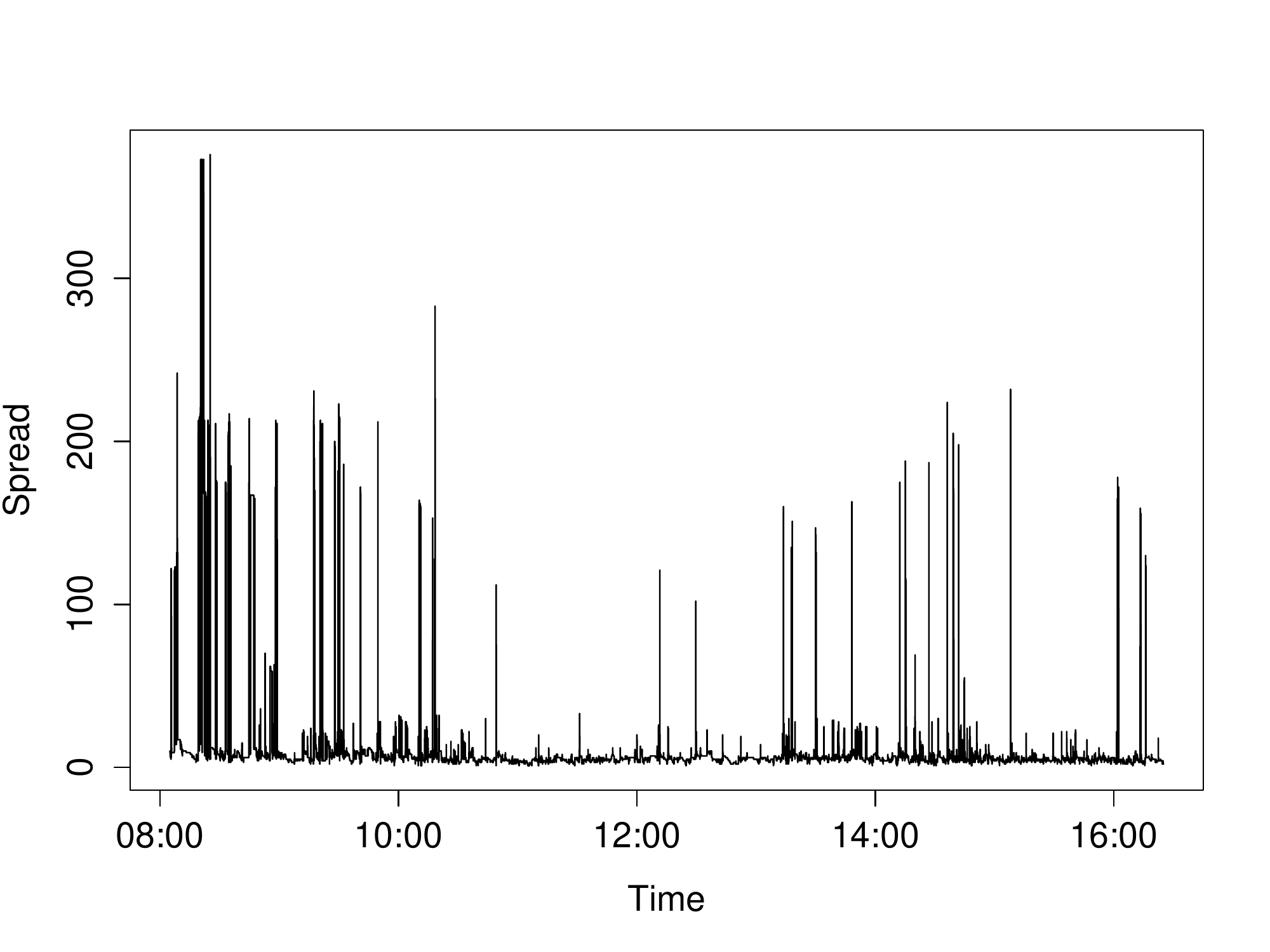}
	\includegraphics[height = 3.5cm,width=0.4\textwidth]{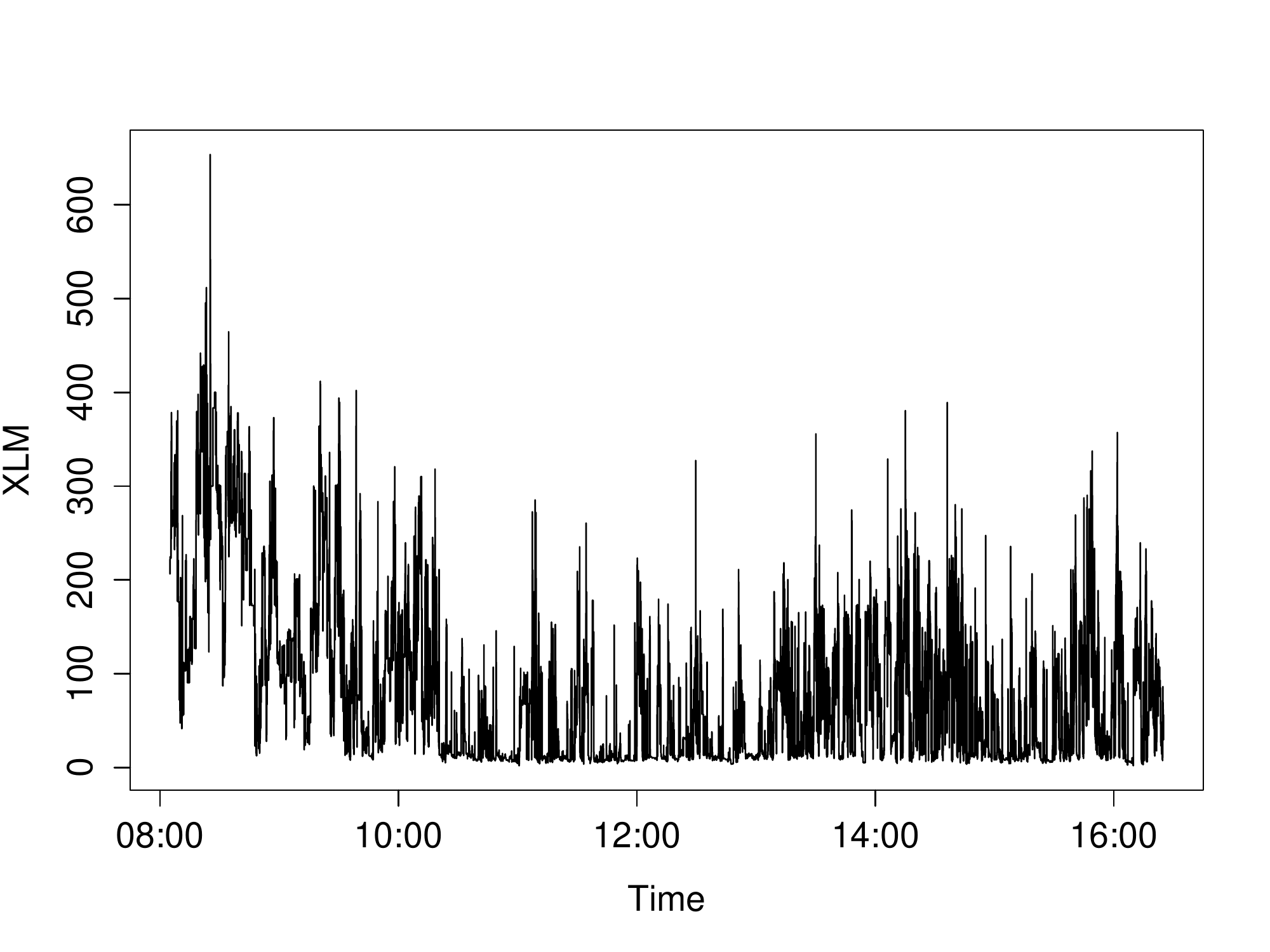}
	\includegraphics[height = 3.5cm,width=0.4\textwidth]{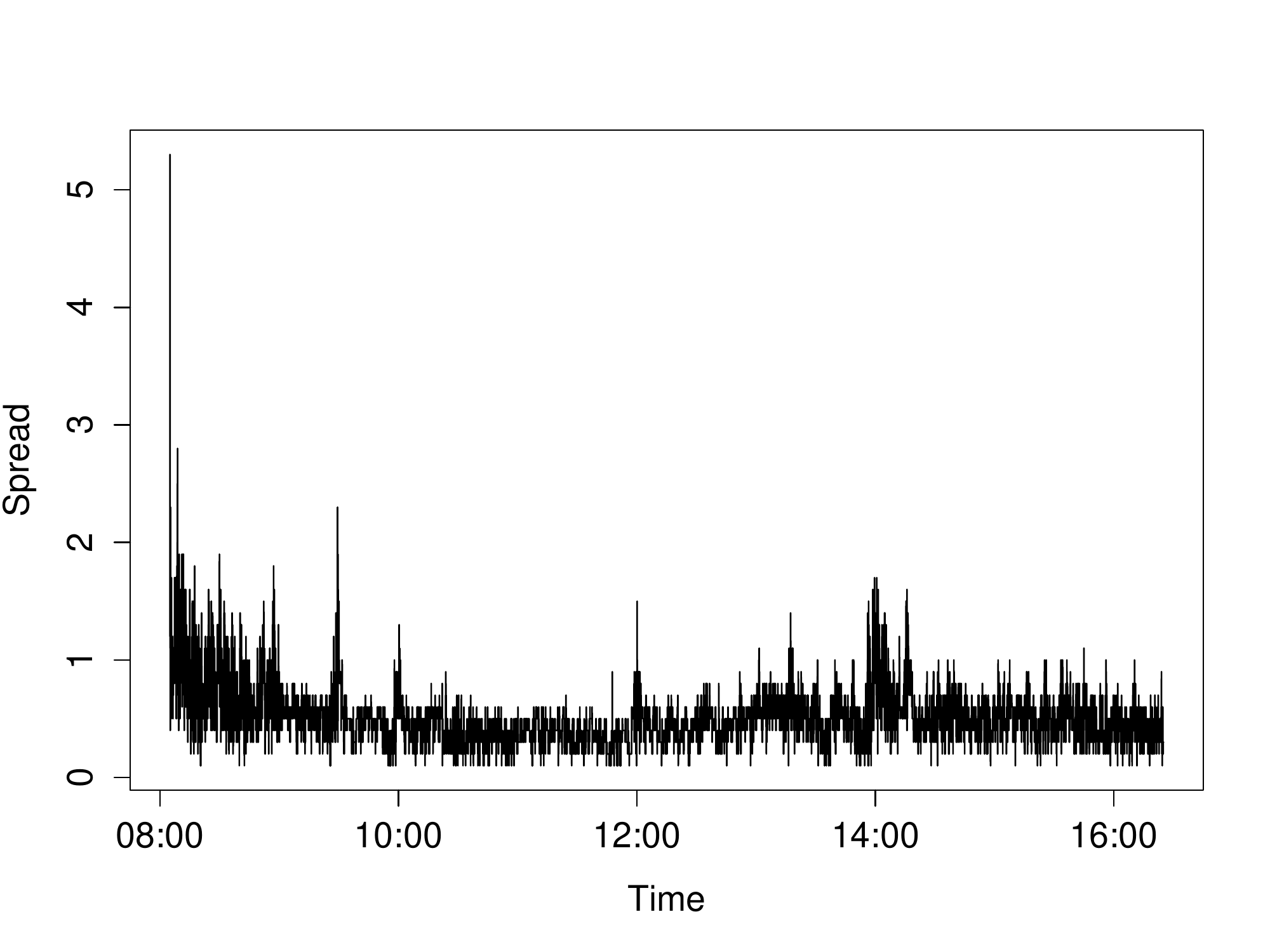}
	\includegraphics[height = 3.5cm,width=0.4\textwidth]{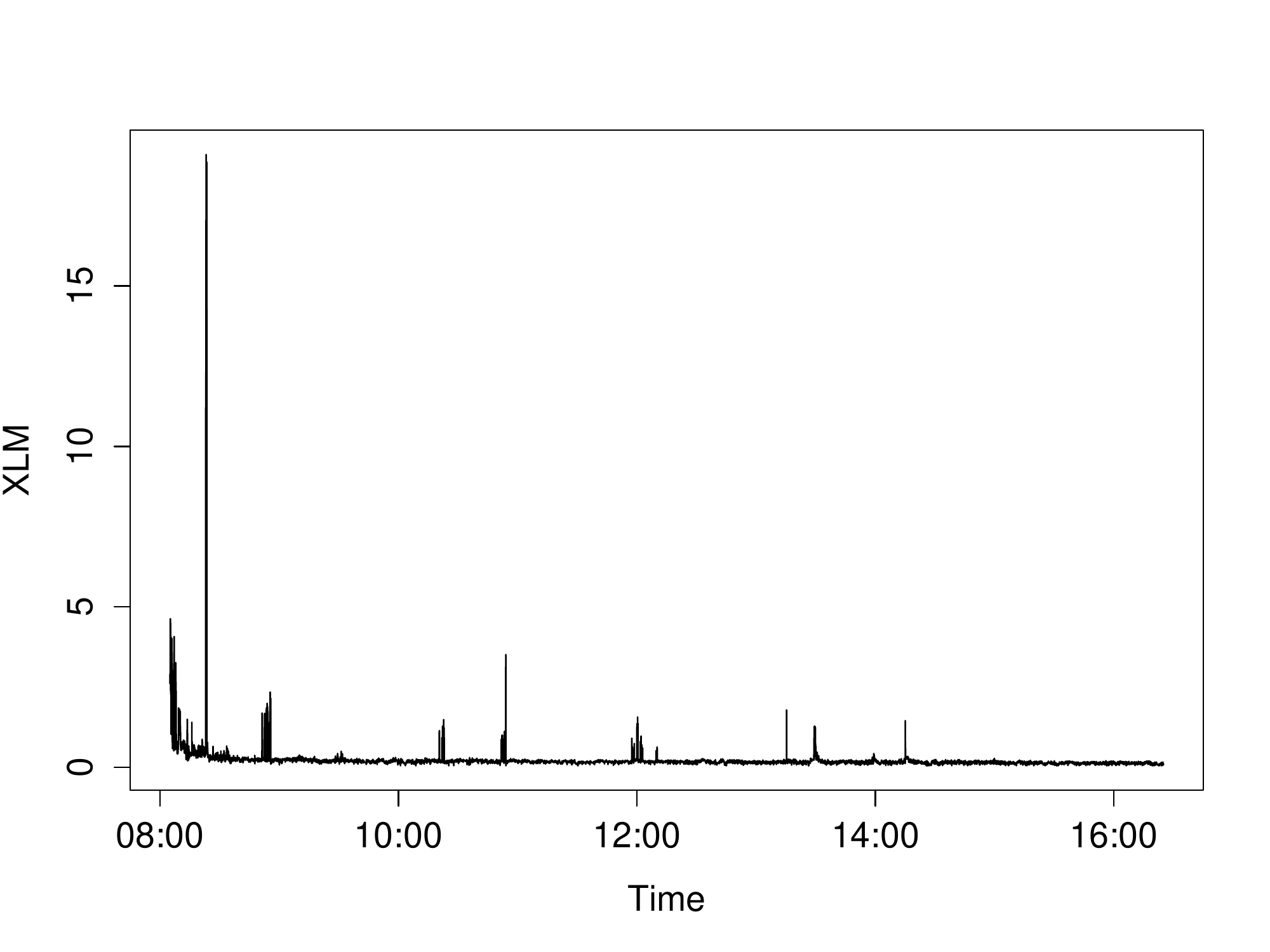}
	\caption{The daily evolution of the spread (left) and XLM (right)  Nexans SA (symbol NEXp, top) and Barclays (symbol BARCl, bottom) on the 15th of February 2012 }
	\label{fig:spread}
	\end{center}
\end{figure}

To further investigate this feature, in Figure \ref{fig:spread} we present a plot of the spread and XLM on a randomly selected day for one of the assets with high $R^2$ (Nexans SA, stock symbol NEXp) and contrast it with the same liquidity measures for a second randomly selected stock with low PCA regression coefficient of determination (Barlcays - stock symbol BARCl). When contrasting these liquidity profiles, we note the very distinct spikes for NEXp, in both liquidity measures, indicating a heavy tailed distribution for these liquidity measures for this asset. Heavy tailed features of liquidity were also observed in the other assets which had exceptionally high $R^2$ values in the PCA regression. 

If one performs the standard PCA approach to extract market factors affecting liquidity, one would see that due to some relatively illiquid periods in the day for certain assets, these assets will dominate the PCA decomposition. This would therefore give a misleading picture of the market contribution to liquidity. If the PCs were then used in a regression, one would then expect that the explanatory power of the PCs for the assets which did not feature such illiquidity spikes would be fairly low, as is the case here. 

Removing the two assets which regularly appear to drive the commonality does not solve the issue, as other assets which contain heavy tailed features then become more prominent. In general, we find a large number of assets whose liquidity measures, to varying degrees, are heavy tailed. We note that these features were not reported in the work of \cite{mancini2013liquidity}, although liquidity in the foreign exchange markets which they investigate is generally much higher than in the equity markets, and therefore such heavy tails may not feature in the distributions of the liquidity measures they consider. Heavy tailed distributions in LOB depth are not specific to equities, however, as they have been identified and studied in \cite{richards2012heavy} also. 

In light of these results, we would suggest that the explanatory power of PCs extracted through liquidity data with heavy tailed features may instead be interpreted as the degree of illiquidity commonality. This is because the leading PCs do not accurately reflect market liquidity, but rather the liquidity of the most illiquid assets. Our analysis reveals that the liquidity for most assets is poorly explained by these illiquid assets, and that there may therefore be, perhaps indirectly, a commonality in the liquidity of the remaining assets. However, this remains to still be studied. This is a different explanatory mechanism for the observed liquidity commonality features in the asset cross-section, compared to those discussed in \cite{chordia2000commonality}, \cite{domowitz2005liquidity} and \cite{hasbrouck2001common}.

One important aspect of our analysis is to highlight the importance of considering the appropriateness of the statistical techniques for large-scale datasets before routine application, see discussion on such matters in the PCA context in \cite{candes2011robust}. We also suggest that the summary statistic or measure one selects for the datasets, in this case the liquidity measure, should be chosen appropriately so as to satisfy the assumptions of the statistical analysis being performed to assess commonality. Based on these findings we argue that it would be pertinent to therefore either consider alternative liquidity measures that don't demonstrate these statistical heavy tailed features so that PCA regressions may still be applied accurately, or to modify the approach adopted for the PCA to account for heavy tailed data, such as via the techniques discussed in \cite{chen2009robust} and \cite{RobustPCAHT}. As a third and perhaps more appropriate alternative, particularly for these large-scale high-frequency LOB liquidity measure datasets exhibiting marginal heavy tailed features, one could instead utilise Independent Component Analysis (ICA). 


\subsection{Independent Component Analysis and Projection Pursuit}
In this section we perform a commonality study of liquidity based on higher order moments, to assess if commonality is still observed in our data having accounted adequately for heavy tailed features of the data.  When liquidity measures in the asset cross-section are either heavy tailed, non-linear in the relationship with a market driving factor, or non-stationary, then one may resort to other forms of decomposition such as ICA, see discussion in \cite{hyvarinen2004independent}.

\begin{landscape}
\begin{figure}[ht!]
  \begin{center}
  \includegraphics[height = 0.3\textheight, width=0.99\linewidth]{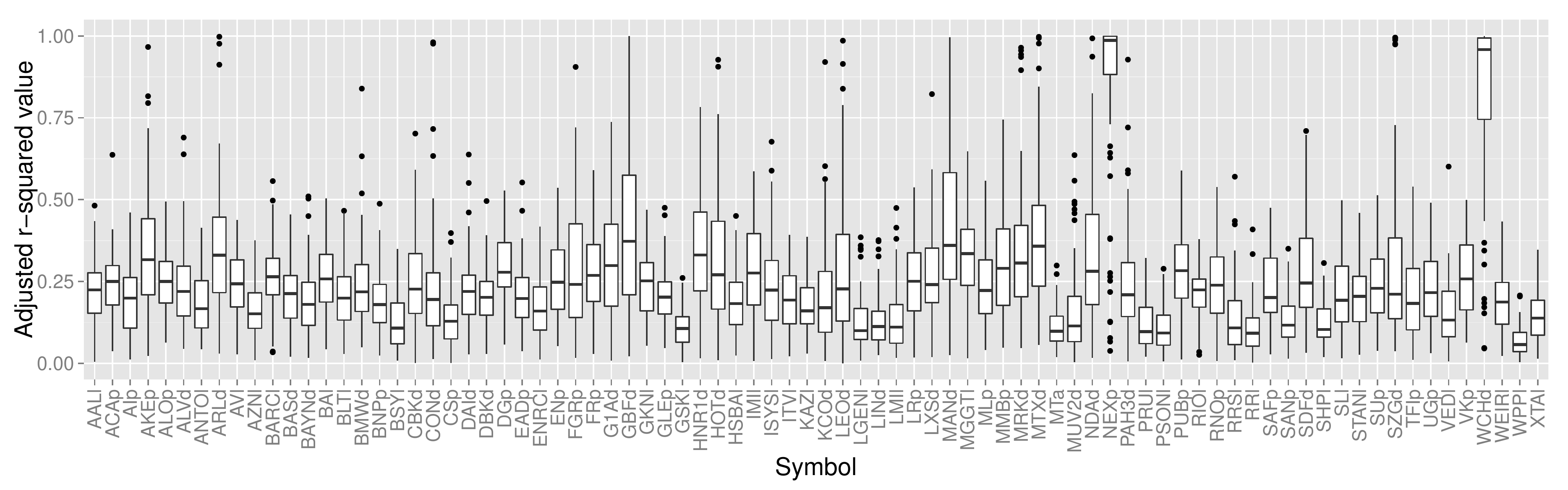}
		\includegraphics[height = 0.3\textheight, width=0.99\linewidth]{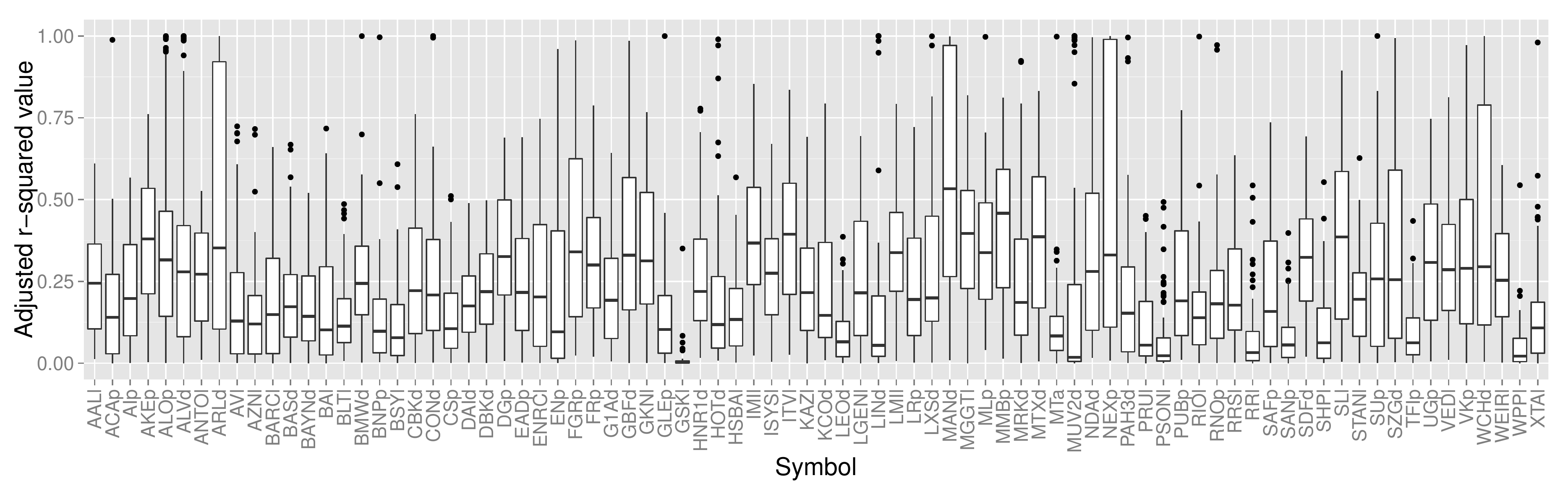}
		\includegraphics[height = 0.3\textheight, width=0.99\linewidth]{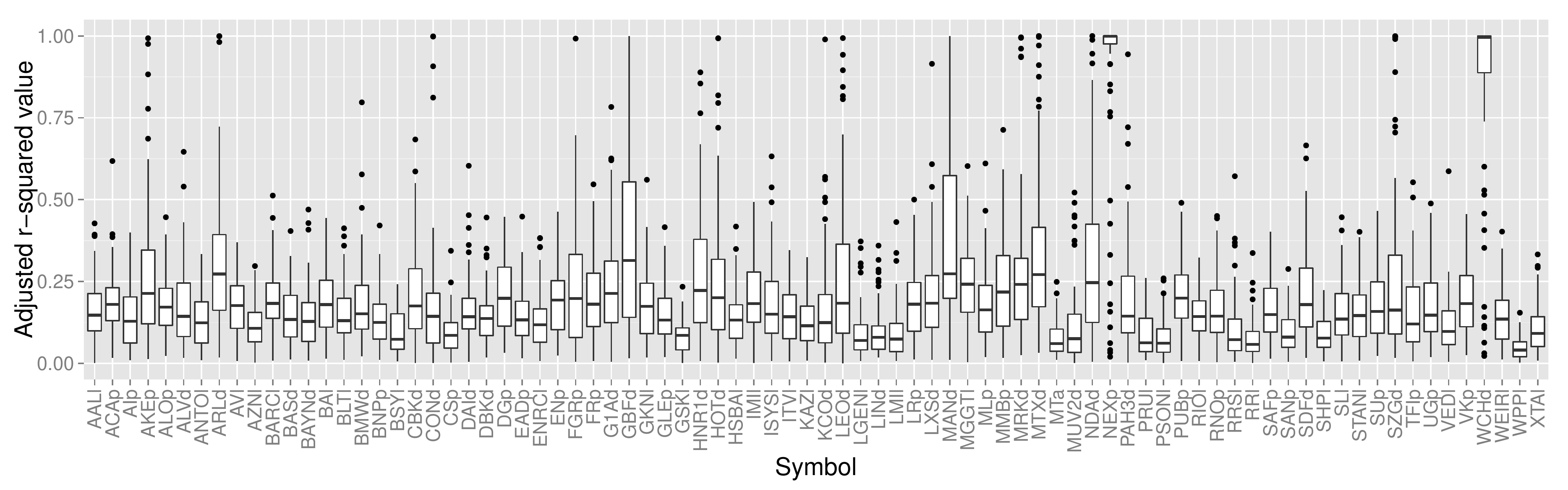}
	\caption{Subplot 1 and 2: The $R^2$ values obtained from regressing individual asset liquidity against the first three PCs obtained across assets for the spread(top) and XLM(bottom). Subplot 3: The $R^2$ values obtained from regressing individual asset liquidity against the first three ICs obtained across assets for the spread}
	\label{fig:allr2pca}
	\end{center}
\end{figure}
\end{landscape}  

ICA methods, in contrast to the correlation-based transformations obtained in PCA for the market liquidity factors, not only de-correlate the liquidity measures in the asset cross-section each day, but also reduce higher-order statistical dependencies. There have been a few studies linking ICA with regression, such as \cite{hyvarinen1997fast}, \cite{shao2006new} and \cite{kwak2008dimensionality}, however this is a relatively under-explored area. Whereas PCA minimises the covariance of the data, ICA minimises higher-order statistics such as the fourth-order cumulant (or kurtosis), thus minimising the mutual information of the output. Specifically, PCA yields orthogonal vectors of high energy content in terms of the variance of the signals, whereas ICA identifies independent components for non-Gaussian signals.



The ICA model equation is an under-determined system and one cannot determine the variances of the independent components. Therefore, one cannot rank the order of dominant components as in PCA, where there is a ranking given by the eigenvalues. In this paper we therefore consider a projection pursuit (PP) based approach, which enables us to select the three leading ICA components, based on a maximization of the negative entropy; see discussion in \cite{girolami1996negentropy}.

After regressing each asset's liquidity against the first three ICA components selected via PP, every day, we again obtained strong evidence to suggest that the assets dominating the PCA analysis due to Gaussianity violations, such as Nexans SA, also correspond to those that were very well explained in a linear projection by the ICA components. The coefficients of determination for the daily regressions for the other assets in the analysis are displayed in boxplots in Subplot 3 of Figure \ref{fig:allr2pca}. In this case, after accounting for higher order statistical features of the cross-section of asset liquidity profiles, if individual assets have weak coefficients of determination with the independent components then this suggests that in a linear structure, these assets display liquidity commonality.

\section{Liquidity resilience for high frequency data}
\label{sec:liqres}
In addition to the considerations regarding the appropriateness of the statistical assumptions for an analysis of commonality, we note that existing liquidity commonality approaches only reflect the aspects of liquidity measure chosen. In the case of the spread, this would be the tightness, and in the case of the XLM, it would also reflect the depth. However, since such measures cannot quantify liquidity resilience (which can be understood as the speed of liquidity replenishment), the associated commonality analysis will not reflect this aspect of liquidity either. Here, we extend the analysis to determine if the liquidity commonality observed is also present when one incorporates notions of resilience.

We employ the resilience notion of \cite{panayi2014market}, which was based on the idea of the Threshold Exceedance Duration (TED):

\begin{definition}
The threshold exceedance duration (TED) is the length of time between the point at which a liquidity measure deviates from a threshold liquidity level (in the direction of less liquidity), and the first point in time at which it returns to at least that level again. 
\end{definition}

In Appendix \ref{sec:TED}, we explain how one can use a parametric survival regression model to model the variation in the TEDs over time, where the duration variable of interest is denoted by $\tau^{TED}$. Using this formulation, one can obtain model based estimates of the expected (log) duration of an exceedance over a chosen threshold $j$ of the liquidity measure, i.e. $E[ln(\tau^{TED,[j]})|\boldsymbol{\hat{\beta}},\mathbf{x}]$ given selected covariates $\mathbf{x}$ that are based on the LOB structure, explained in Appendix \ref{sec:TED} and in detail in \cite{panayi2014market}.

\begin{figure}[ht!]
	\begin{center}
	\includegraphics[width=0.49\textwidth]{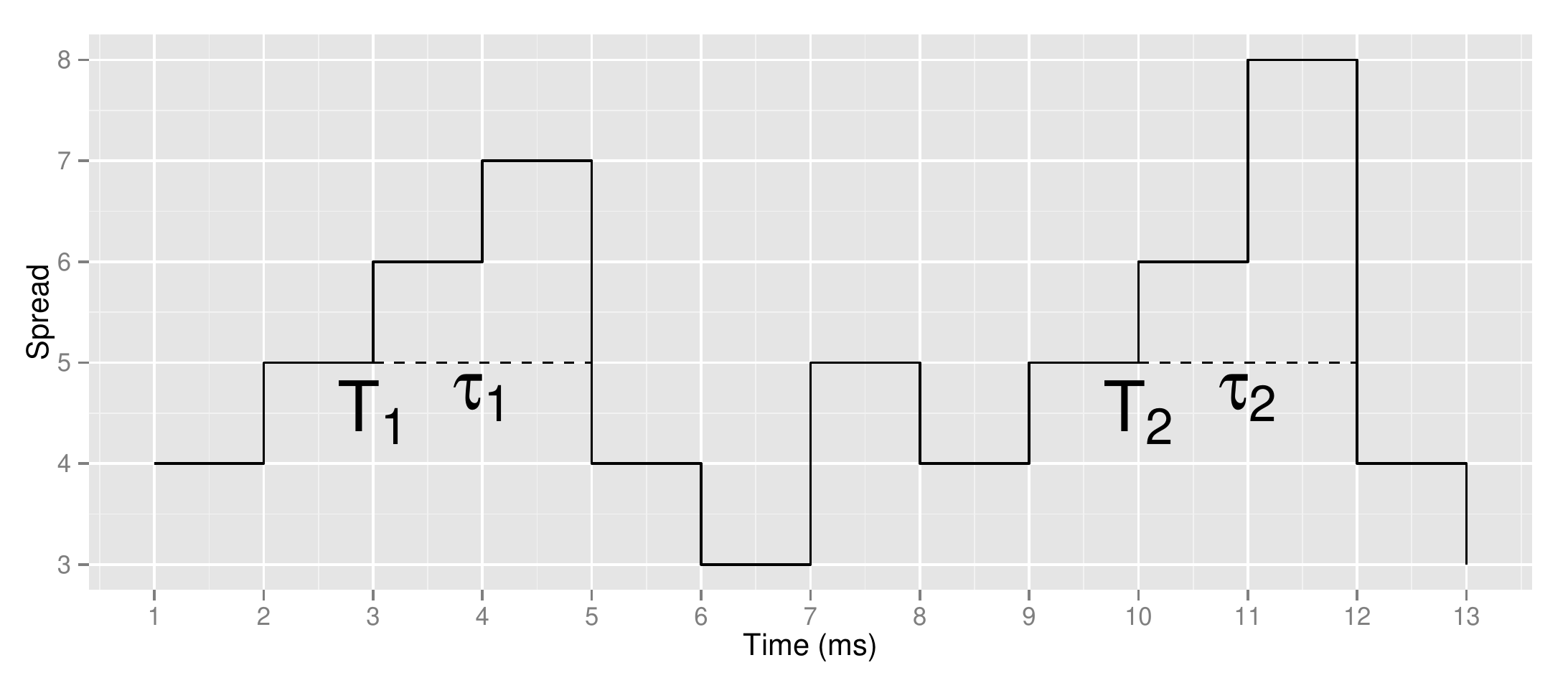}
	\caption{An example of two threshold exceedance deviations, where the liquidity measure used in the TED definition is the spread, and the threshold is 5 ticks.}
	\label{fig:TED}
	\end{center}
\end{figure}

Alternative notions of liquidity resilience have been suggested by \cite{kyle1985continuous}, as the speed with which prices recover from a random, uninformative shock, or by \cite{foucault2005limit}, as the probability of the spread reverting back to its former level before the next trade. The TED metric is more general, however, as it allows one to select both the liquidity measure and the threshold liquidity level of interest. For example, a brokerage firm executing an algorithm would be interested in deviations from a high liquidity threshold level to minimise execution costs, while a regulator might be interested in exceedances below a critically low threshold level, as part of its review into market quality.     

\subsection{Summarising resilience behaviour}
Under the TED formulation of \cite{panayi2014market}, $E[ln(\tau^{TED,[j]})|\boldsymbol{\hat{\beta}},\mathbf{x}]$ can be obtained for a large range of liquidity thresholds which can be combined to obtain the \textit{Liquidity Resilience Profile} (LRP). The LRP is a summary of the expected resilience behaviour of an asset across different liquidity thresholds:

\begin{definition}
The daily \textit{Liquidity Resilience Profile} is a curve of the expected TEDs as a function of the liquidity threshold, given the state of the LOB, as quantified by the covariates characterising the LOB for the given asset.
\end{definition}

\begin{figure}[ht!]
	\begin{center}
	\includegraphics[height = 3.5cm,width=0.45\textwidth]{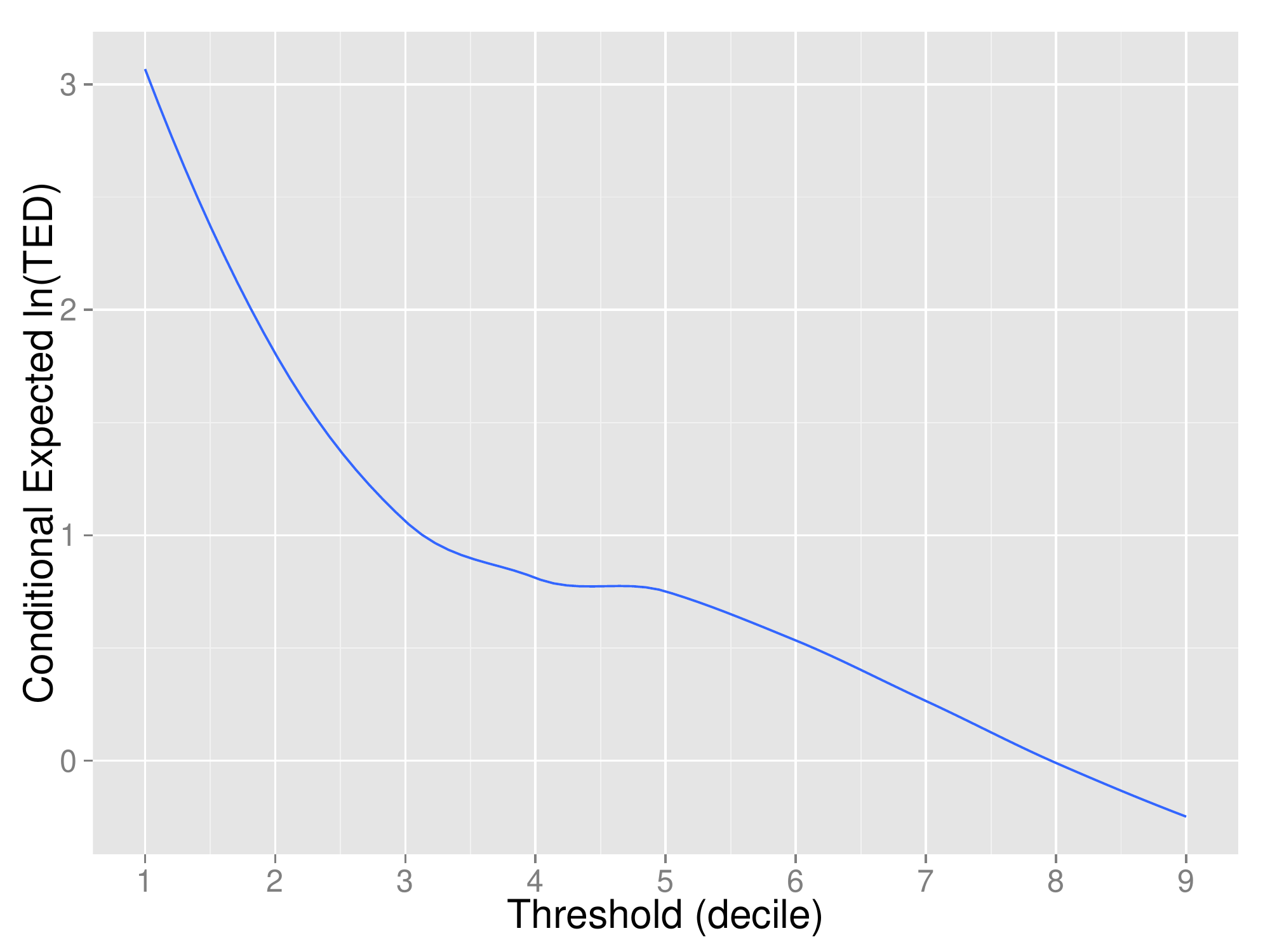}
		\includegraphics[height = 3.5cm,width=0.45\textwidth]{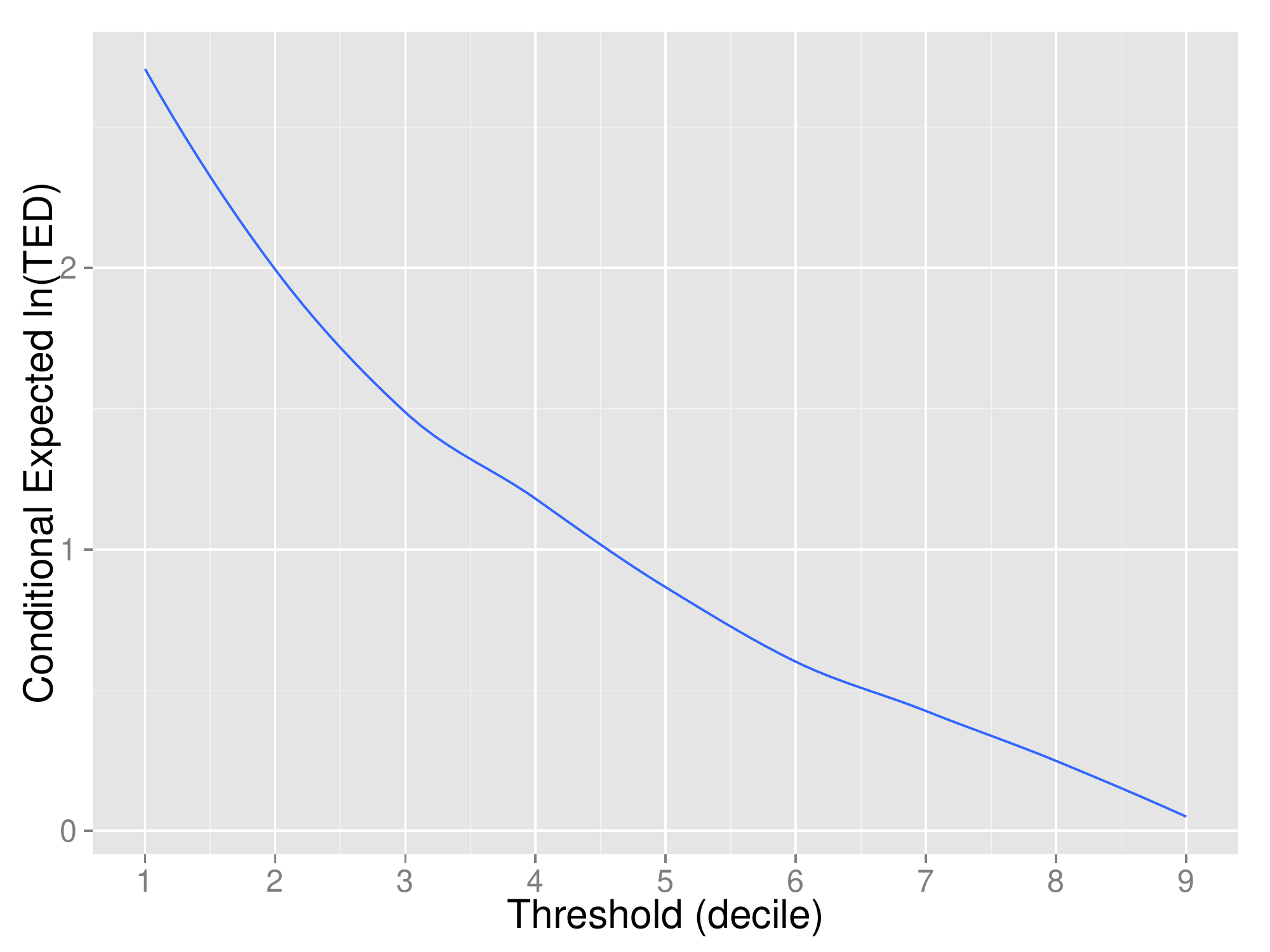}
	\caption{Liquidity resilience profile for Credit Agricole in the normal LOB regime, in which covariates take their median values. The x-axis represents the threshold used in the TED definition, and the curve is obtained by considering thresholds corresponding to deciles of the empirical distribution of the liquidity measure - in this case, the spread(left) and XLM(right).}
	\label{fig:liqresACAp}
	\end{center}
\end{figure}

To facilitate comparison between the liquidity resilience behaviour of assets at different threshold levels, we present results for the logarithm of the expected TED, i.e. we calculate $E[ln(\tau^{TED,[j]})|\boldsymbol{\hat{\beta}}^{[j]},\mathbf{x}^{[j]}]$, where the $j$ are threshold levels $j=1,\ldots,9$ corresponding to deciles of the empirical spread or XLM distribution. We can then identify the commonality in the expected TED over the median spread, or the 9th decile of the XLM, for example. We explain how the smooth functional representation is obtained in the following section. 

We should note here that we will diverge somewhat from previous analyses, which only quantified the \textit{temporal} commonality between the liquidity of individual assets. The temporal component of liquidity resilience commonality is captured through the similarity in the expected daily exceedance times over a threshold (measured through the TED metric). However, as we have obtained a representation of the expected exceedance time as a function of the level of the threshold, we can also quantify this commonality at different thresholds. 

The functional representation enables us to then establish whether such commonality in liquidity resilience behaviour exists at any or all levels of the liquidity measure. The part of the curve corresponding to low thresholds of the spread or XLM indicates the expected time to return to a high level of liquidity, which would interest a brokerage house trying to minimise execution costs. The part of the curve corresponding to high thresholds, on the other hand, indicates the expected duration of periods with very low liquidity, which could be considered by a regulator as part of their efforts to ensure uninterrupted liquidity in financial markets.  

\section{Functional data analysis characterisations of massive LOB data sets}
\label{sec:fdaintro}
In the analysis of financial data, one often has to deal with the issue of the high dimensionality of the data sets under consideration. We argue that there are significant advantages in summarising such high dimensional massive data sets under a functional characterisation. Functional data analysis (FDA) is a statistical approach that can be used to reduce the dimensionality of the problem, like a PCA analysis, but it allows one to capture additional features and perform analysis that are not possible in standard PCA approaches in the Euclidean space. 

FDA has several advantages compared to multivariate analysis \cite{coffey2011common}. It can achieve a parsimonious representation of an entire series of measurements from a single source, as a single functional entity, rather than a set of discrete values. It can also account for the ordering of the data (time-based or otherwise) through smoothing, as it is unlikely that adjacent values will vary by a large amount \cite{ramsay2006functional}. In addition, compared to multivariate analysis, it does not require that concurrent measurements are taken from every source of information. For these reasons, FDA would be highly appropriate for the analysis of unevenly spaced and high dimensional financial data. Detailed accounts of each aspect of FDA are provided in the text of  \cite{ramsay2006functional}.

Once a functional representation is obtained, one can explore functional equivalents of analyses performed in the multivariate space. For example, one can perform functional principal components analysis (FPCA) to extract the leading \textit{eigenfunctions} characterising the functional dataset. Canonical correlation analysis can also be applied in the functional space, in order to investigate the modes of variability in two sets of functions that are most associated with one another. There are different ways in which one can build a functional linear model, with either a functional dependent variable, a set of functional covariates, or both \cite{ramsay2006functional}. We use the \textit{concurrent model} in this paper, involving a form of functions on functions regression, where we assume that the response is only affected by the dependent variables at the same point of the domain of the functions.


In the next section we detail how one can use FDA to obtain functional representations of LRPs, and FPCA to extract the dominant modes of variation every day for the asset cross-section. In addition, we will build a concurrent functional multiple regression model to quantify the explanatory power of the functional principal components (FPCs) for the LRPs of individual assets.     

\section{Functional data summaries: smoothed functional representations for LRPs}
\label{sec:lrpfits}

Functional data analysis is the study of functional data, where the domain of the function is usually time, but could be frequency, space, or in our case, thresholds of a liquidity measure. It differs from multivariate analysis methods such as time series modelling in its imposition of smoothness constraints:
\begin{equation}
\mathbf{y}=x(\mathbf{u})+\boldsymbol{\epsilon}
\end{equation}
where the $x(u)$ is considered to be a smooth functional data observed at certain points $\mathbf{u}=(u_{1},\ldots,u_{n})$, in the presence of noise, to get observations $\mathbf{y}=(y_{1},\ldots,y_{n})$. FDA then enables us to describe the variation in functional data, obtain derivatives and cluster curves according to their similarity. 

In this paper our interest is in the expected liquidity resilience behaviour of every asset for different thresholds. In this context, the dependent $y_j$ is $E[ln(\tau^{TED,[j]})|\boldsymbol{\hat{\beta}}^{[j]},\mathbf{x}^{[j]}]$, where the $j$ are threshold levels $j=1,\ldots,9$, defined as deciles of the empirical distribution of the liquidity measure. We will first explain how to obtain a smoothed representation $x(u)$ of the liquidity resilience profile of every asset, and then determine whether the dominant modes of variation over the different assets can be explanatory for the resilience of individual assets over the long term.

\subsection{Defining a basis system for functional data representation}
The first challenge is obtaining a functional data representation of discrete (and possibly noisy) observations of the daily LRPs for each asset. We can represent the LRP function $x(u)$ on a given day, for a given asset, using a basis expansion method, where a linear combination of the $K$ basis functions $\phi_k(u)$ (with coefficients $c_k$) can approximate a smooth function for a sufficiently large $K$:
\begin{equation}
x(u)= \sum_{k=1}^{K} \phi_k(u)c_k
\end{equation} 
If we then have N functions then 
\begin{equation*}
x_i(u) = \mathbf{c}_i^T \boldsymbol{\phi}(u), i=1 \ldots N
\end{equation*}
Common examples of bases used in FDA include Fourier bases, which are useful when data is periodic, and spline bases, of which several may be considered (B-splines, M-splines, I-splines etc.). Splines are piecewise polynomials, taking values in sub-intervals of the observation range. They are defined by:
\begin{itemize}
\item the range $[u_0,u_L]$ in which they take values;
\item the order $m$ of the spline, which is one higher than the highest degree polynomial;    
\item break points and knots, or the points which divide the observation range. Over a particular subinterval, the order of the polynomial is fixed. There can be several knots at a particular break point, if more than one basis function takes values in an adjacent subinterval. 
\end{itemize}
We choose a B-spline basis here, and B-splines are defined recursively from lower order B-splines as follows:
\begin{align*}
B_{i,0}(u)&=\left\{\begin{matrix}
1, u_i \leq u < u_{i+1}\\ 
0, \text{elsewhere}
\end{matrix}\right.\\
B_{i,j+1}(u)&=\alpha_{i,j+1}(u)B_{i,j}(u)+\left [ 1-\alpha_{i,j+1} \right ](u)B_{i+1,j}(u)
\end{align*} 
with $\sum_i B_{i,j}(u)=1$ and
\begin{equation}
\alpha_{i,j}(u)=\left\{\begin{matrix}
\frac{u-u_i}{u_{i+j}-u_i},\text{if } u_{i+j}\neq u_i\\ 
0,\text{otherwise.}
\end{matrix}\right.
\end{equation}
The spline function $S(u)$ is then defined as 
\begin{equation}
S(u)=\sum_{k=1}^{m+L-1}c_k B_{k,m}(u)
\end{equation}

This formulation means that a basis function is positive over at most $m$ subintervals (this is called the compact support property), making estimation efficient. In our application, we obtain the LRP curve as a functional representation of the daily expected conditional log TED at each threshold. 
We achieve this via a cubic B-spline basis (i.e. the order $m=4$).  There is a continuity and smoothness restriction that adjacent polynomials (and first two derivatives) are constrained to be equal at the knots.  In the range of observation thresholds $[u_0,u_L]$ we consider $L-1$ interior knots, the interior knot sequence is generically denoted by $\mathbf{u}=(u_1,\ldots,u_{L-1})$. This produces a total of $m+L-1$ basis functions for the function representation we adopt for the LRP of each asset each day. 

We thus have to select the number of breakpoints (i.e. the value of $L$) and the number of basis functions $K$ to use in the B-spline basis, although selecting a value for one will determine the other. As our curves are constructed over 9 thresholds of the liquidity measure, we select $L=4$ in order to obtain a parsimonious representation of the LRP. 

\subsubsection{Estimation of functional representations of LRPs} 
We perform linear regression on the basis functions to obtain the coefficient vector $\mathbf{c}$, i.e. by minimising the sum of squared errors
\begin{equation}
SSE=\sum_{j=1}^{n} (y_j - \sum_{k=1}^{K} c_k\phi_k(u_j))^2=(\mathbf{y}-\Phi\mathbf{c})'(\mathbf{y}-\Phi\mathbf{c})
\end{equation}
where $\Phi$ is an $N$ by $K$ matrix containing $\phi_j(t_k)$. The OLS estimate is 
\begin{equation}
\hat{\mathbf{c}}=(\Phi^T \Phi)^{-1} \Phi^T \mathbf{y}
\end{equation}
and the vector of fitted values is 
\begin{equation}
\hat{\mathbf{y}}=\Phi \hat{\mathbf{c}}=\Phi (\Phi^T \Phi)^{-1} \Phi^T \mathbf{y}
\end{equation}
from which we can see that $\Phi (\Phi^T \Phi)^{-1} \Phi^T$ acts as a simple linear smoother. This approximation is only appropriate if we assume i.i.d errors, but this is not often the case with functional data. In order to enforce smoothness, we can add a roughness penalty to the least squares criterion
\begin{equation}
PENSSE_{\lambda}=(\mathbf{y}-\Phi\mathbf{c})'(\mathbf{y}-\Phi\mathbf{c}) + \lambda J(x)
\end{equation}
where $\lambda$ is a tuning parameter and $J(x)$ measures roughness, for example through the curvature $J_2(x)=\int_u [D^2 x(u)]^2 du$, or, more generally, using any linear differential operator $J(x)=\int_u \sum_k=1^m \alpha_k D^k[(x(u))] du$. The $D$ operator is used to denote derivatives, such that $D^0 x(u)=x(u)$ and $D^m x(u) = \frac{d^m x(u)}{du}$.

We impose the $J_2$ roughness penalty in the estimation, as in areas where the function is highly variable, the square of the second derivative will be large. A theorem from \cite{de2001calculation} shows that when choosing $J_2(x)=\int_u [D^2 x(u)]^2 du$, a cubic spline with knots at points $u_j$ minimises $PENSSE_{\lambda}$.  Spline smoothing with the roughness penalty above is still a linear operation, where the smoother is now $(\Phi^T \Phi + \lambda \mathbf{R})^{-1} \Phi^T$, where
\begin{equation}
\mathbf{R}=\int D^2 \boldsymbol{\phi}'(u) D^2 \boldsymbol{\phi}(u)du
\end{equation}
see \cite{ramsay2006functional} for a derivation. This is usually computed by numerical integration.

\begin{figure}[ht!]
	\begin{center}
		\includegraphics[width=0.45\textwidth]{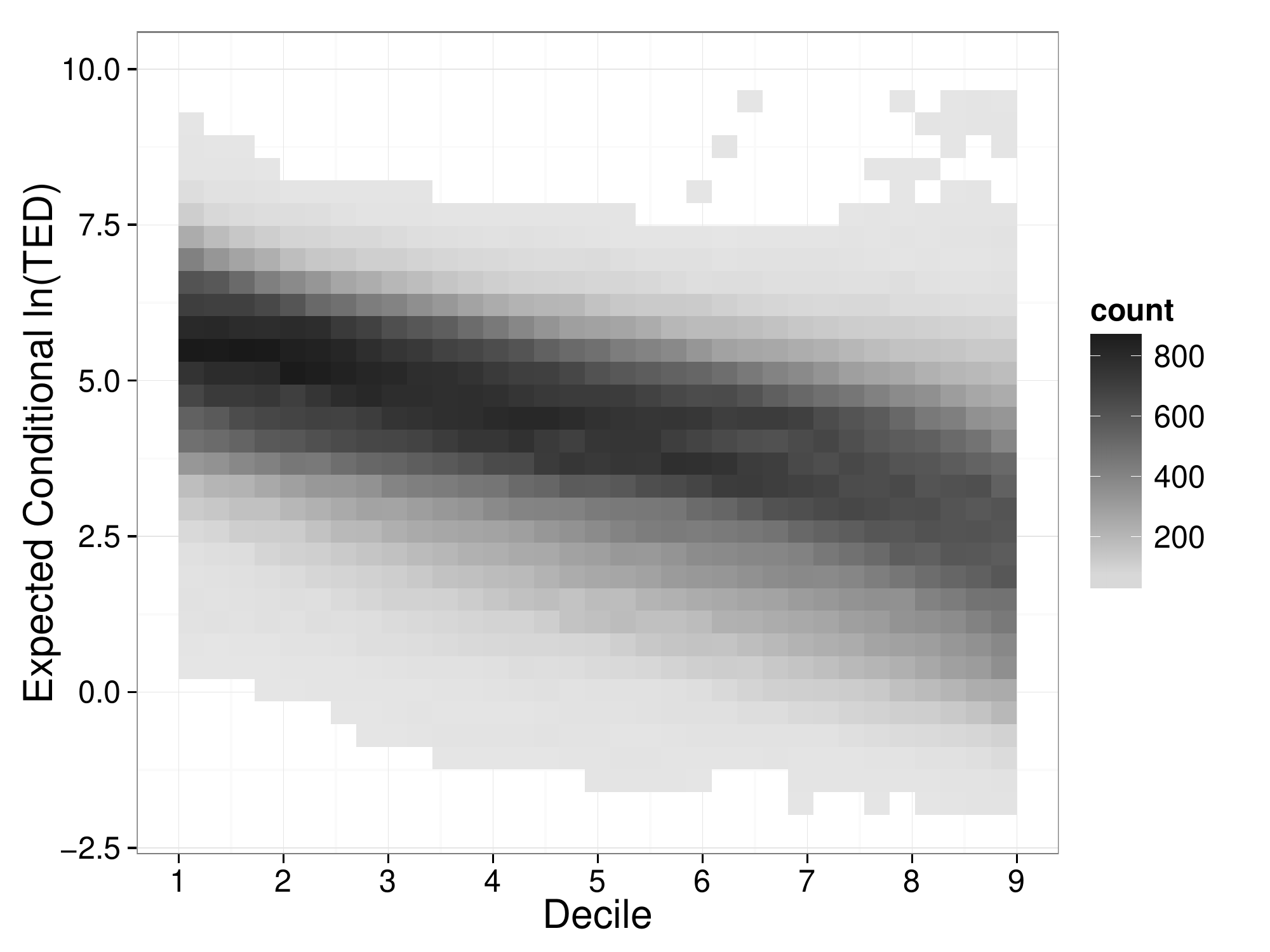}
	\includegraphics[width=0.45\textwidth]{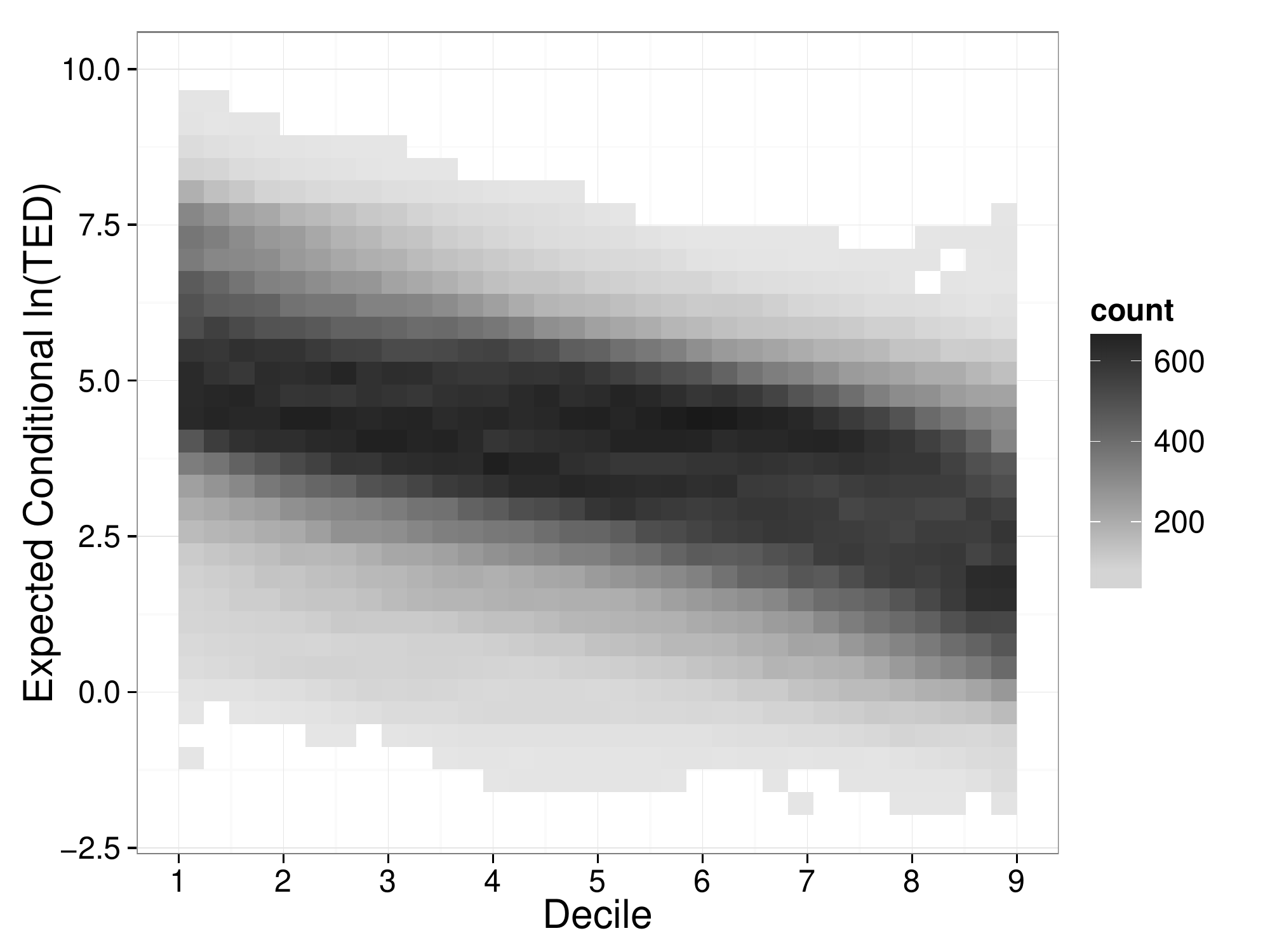}
	\caption{Projected LRPs of all 82 assets, for the entire 81 day period, onto a single grid for the spread (left) and XLM (right).}
	\label{fig:allLRP}
	\end{center}
\end{figure}

The last point is choosing the smoothing parameter $\lambda$, and a widely used approach is the generalised cross-validation (GCV) method proposed by \cite{craven1978smoothing}. One can use the generalised cross-validation measure, whereby minimisation of the criterion is a method to select $\lambda$. For each asset, on every day, we calculate the GCV value on a fine grid, and present in Figure \ref{fig:gcv} in the  histogram of the values of $\lambda$ corresponding to the lowest GCV. We find that these values concentrate at very low levels and very high levels. However, empirical analysis shows that using a large smoothing parameter (close to 1) leads to oversmoothing, and we lose some of the interesting features in the data. For this reason, we choose to use the same parameter $\lambda=0.02$ for every asset, on every day. We summarise the result of this data preparation in Figure \ref{fig:allLRP} for the LRPs for all assets for both the spread and the XLM, obtained using the B-spline basis and roughness penalisation method described above. 
There is variation in the LRPs of individual assets over time, and this justifies our choice of investigating liquidity resilience (and its commonality) daily first. The darker shaded area shows that there is a clustering of LRPs across assets and across time, and this is a first visual confirmation of commonality in liquidity resilience over the different (relative) liquidity measure thresholds. Now that we have estimated smoothed representations of the daily LRPs for every asset, we will treat them as the observed data that we will analyse to quantify any commonality. 

\section{Functional principal components analysis}
\label{sec:fpca}
We now evaluate the market factors characterising the asset cross-section functional LRP profiles each day. In functional principal components analysis (FPCA), we seek the dominant modes of variation over a set of curves. As an example of PCA in the multivariate space, \cite{hasbrouck2001common} and \cite{korajczyk2008pricing} characterise market-wide liquidity as the first principal component of individual FX rates. We focus on the functional equivalent, but want to characterise the resilience of liquidity, rather than liquidity itself. We then determine the extent to which these can be explanatory for individual asset resilience over time.
\begin{figure}[ht!]
	\begin{center}
		\includegraphics[width=0.32\textwidth]{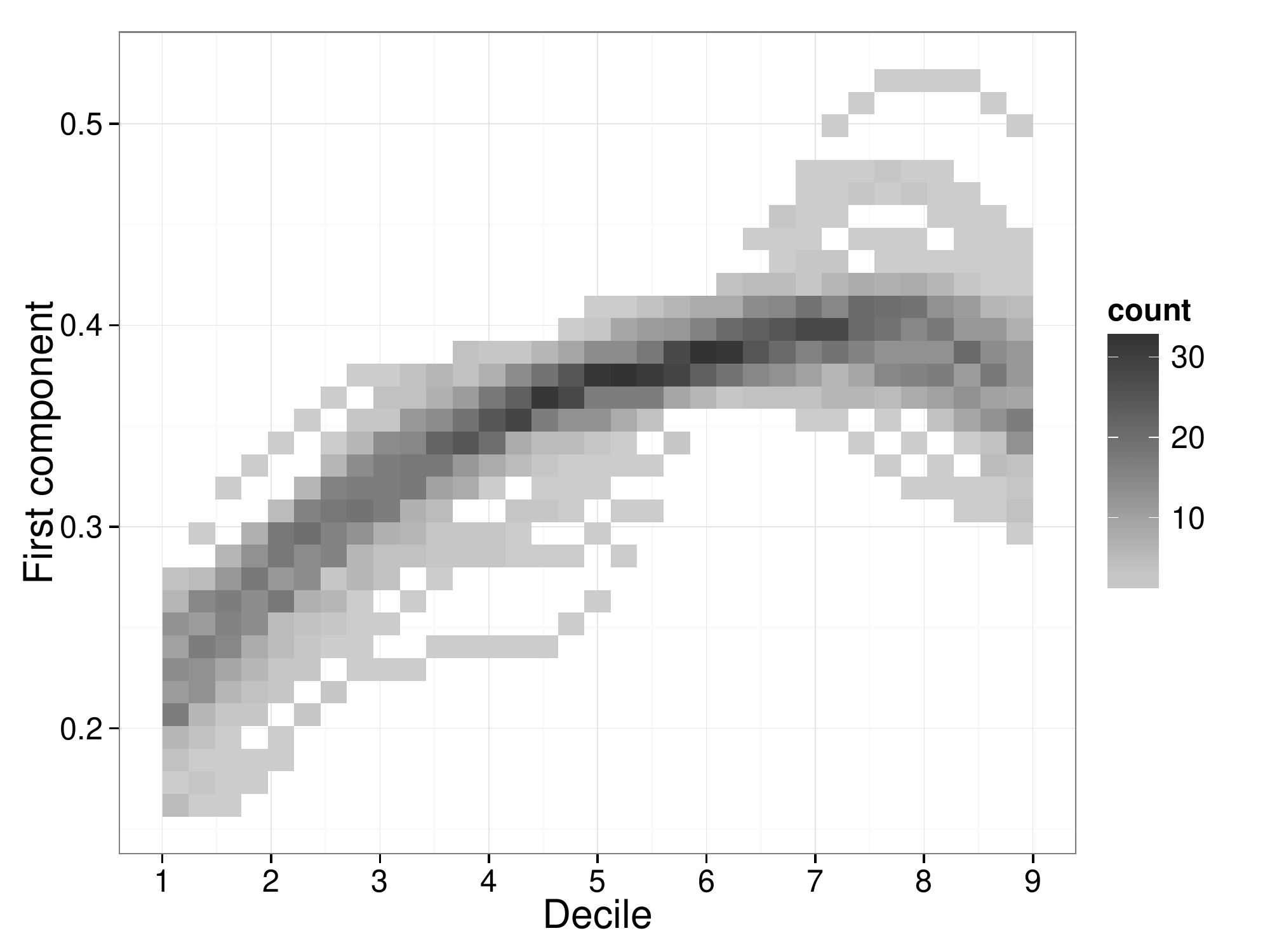}
	\includegraphics[width=0.32\textwidth]{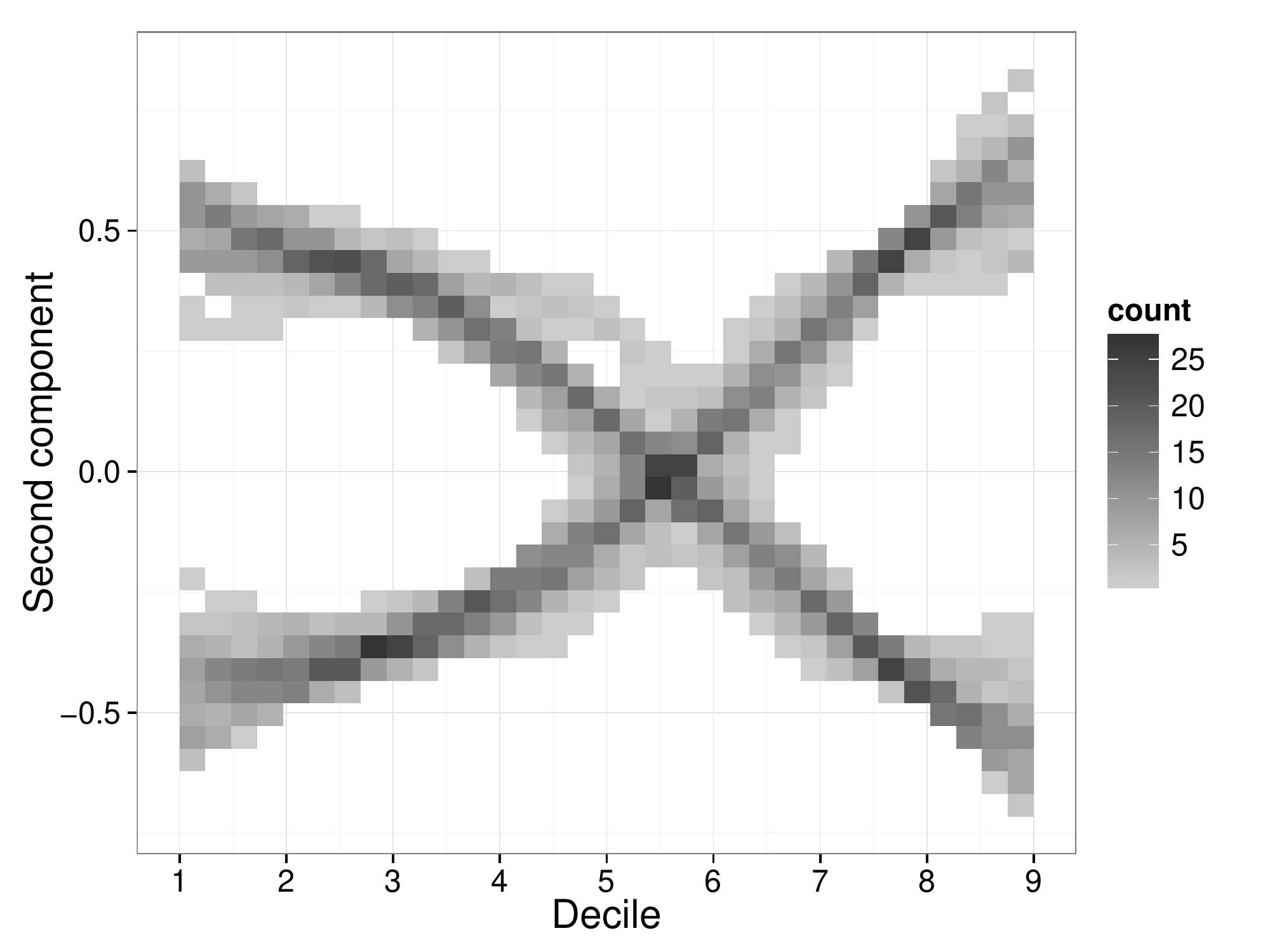}
		\includegraphics[width=0.32\textwidth]{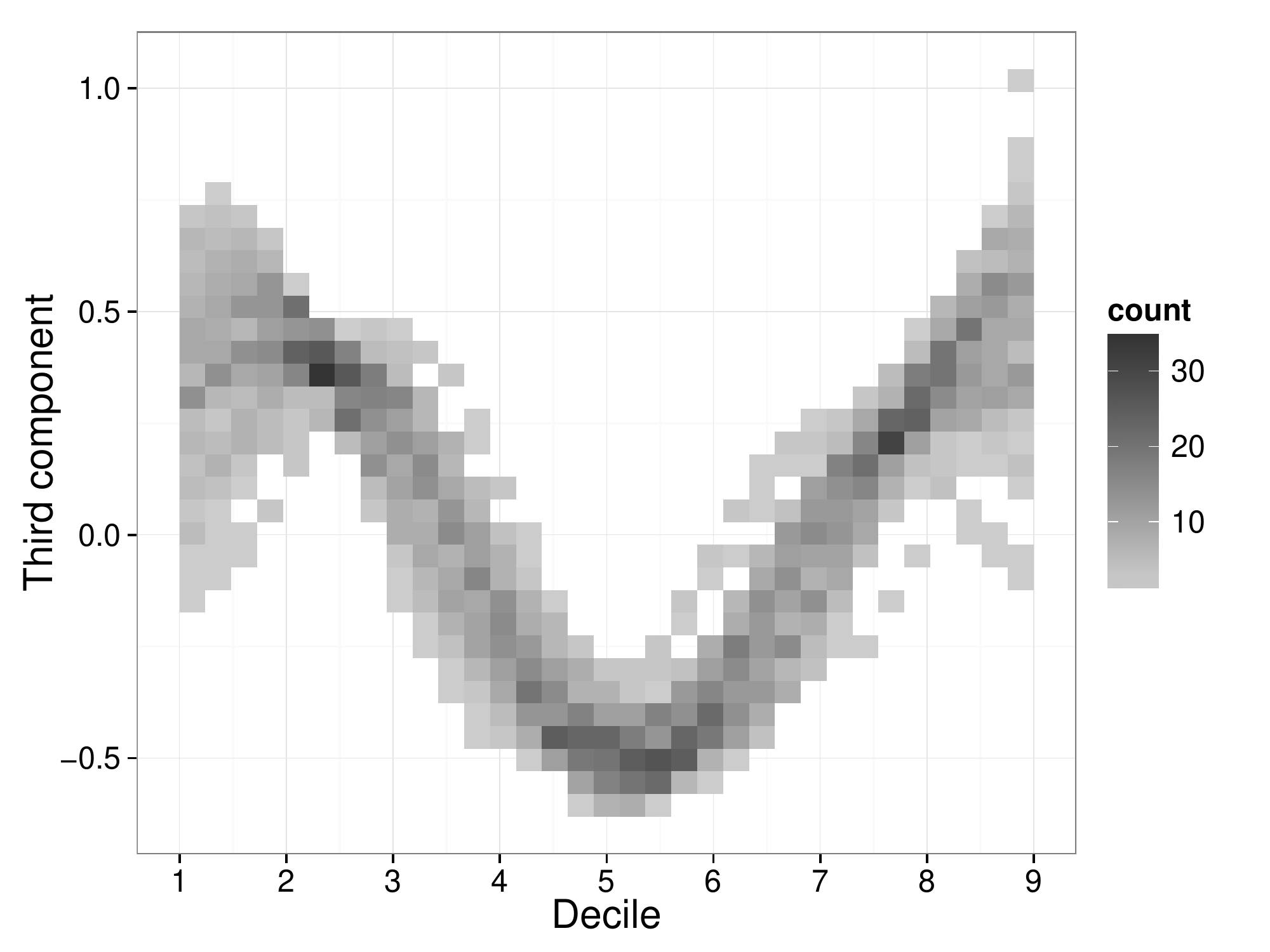}
		\includegraphics[width=0.32\textwidth]{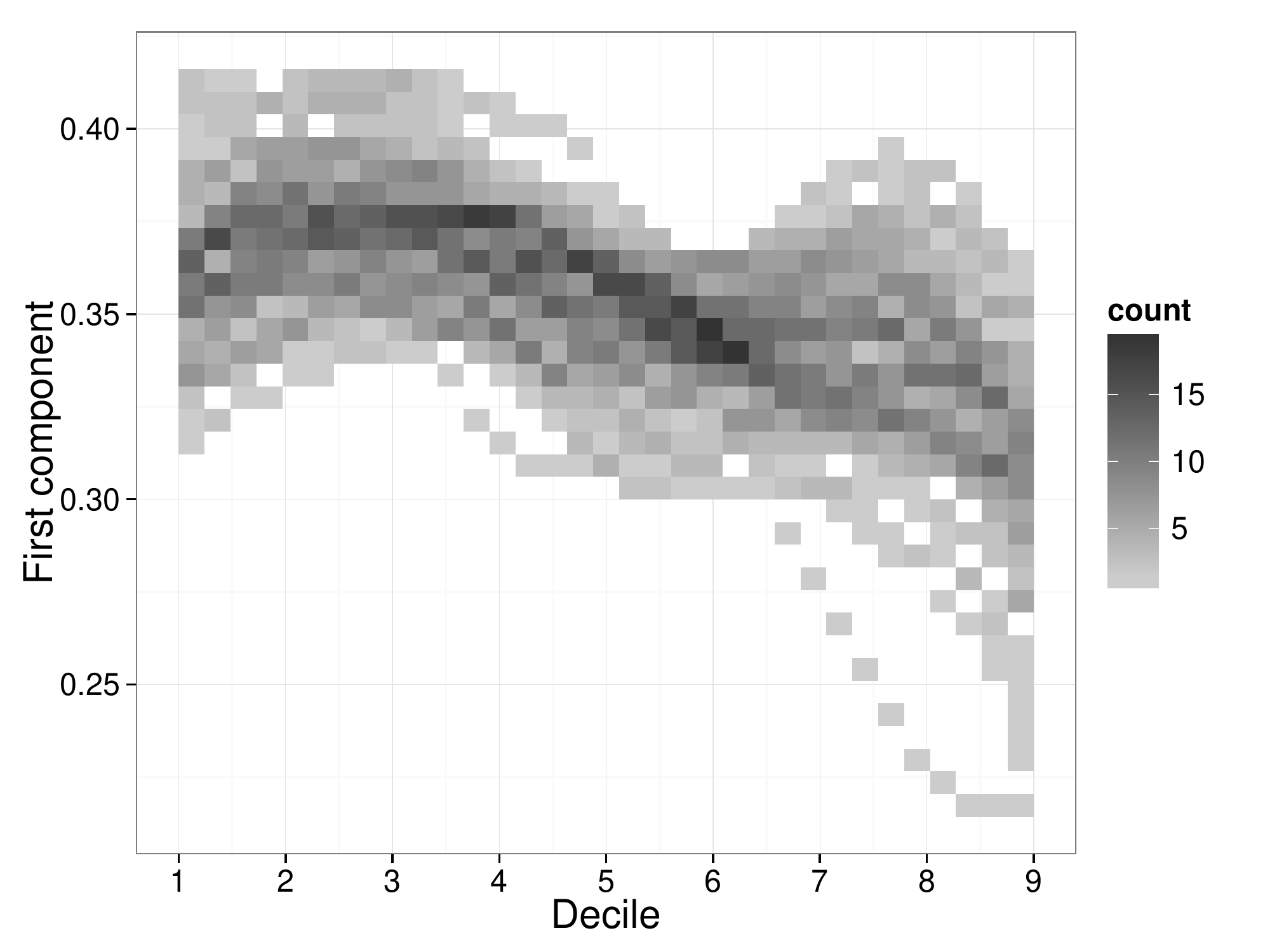}
	\includegraphics[width=0.32\textwidth]{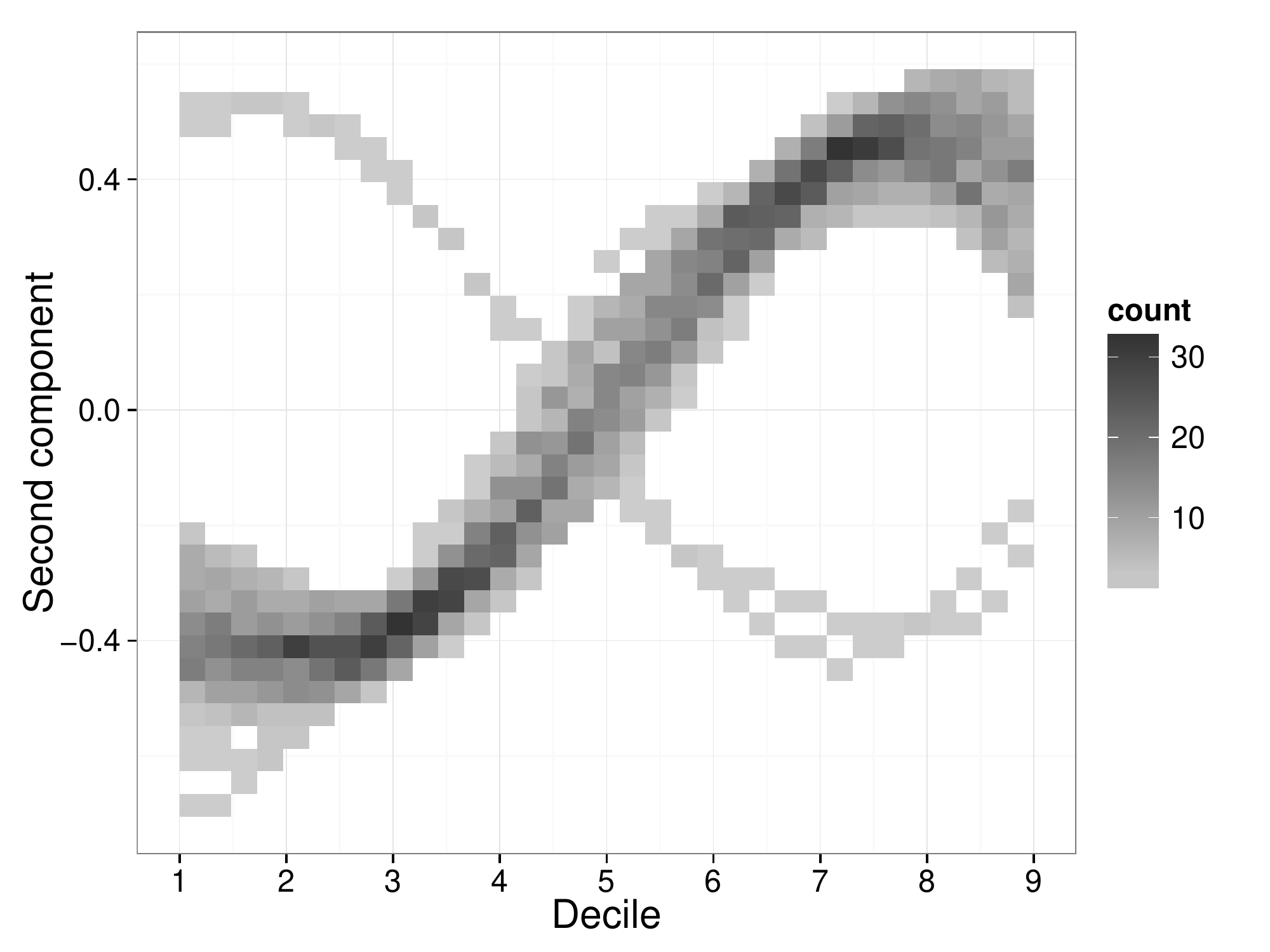}
		\includegraphics[width=0.32\textwidth]{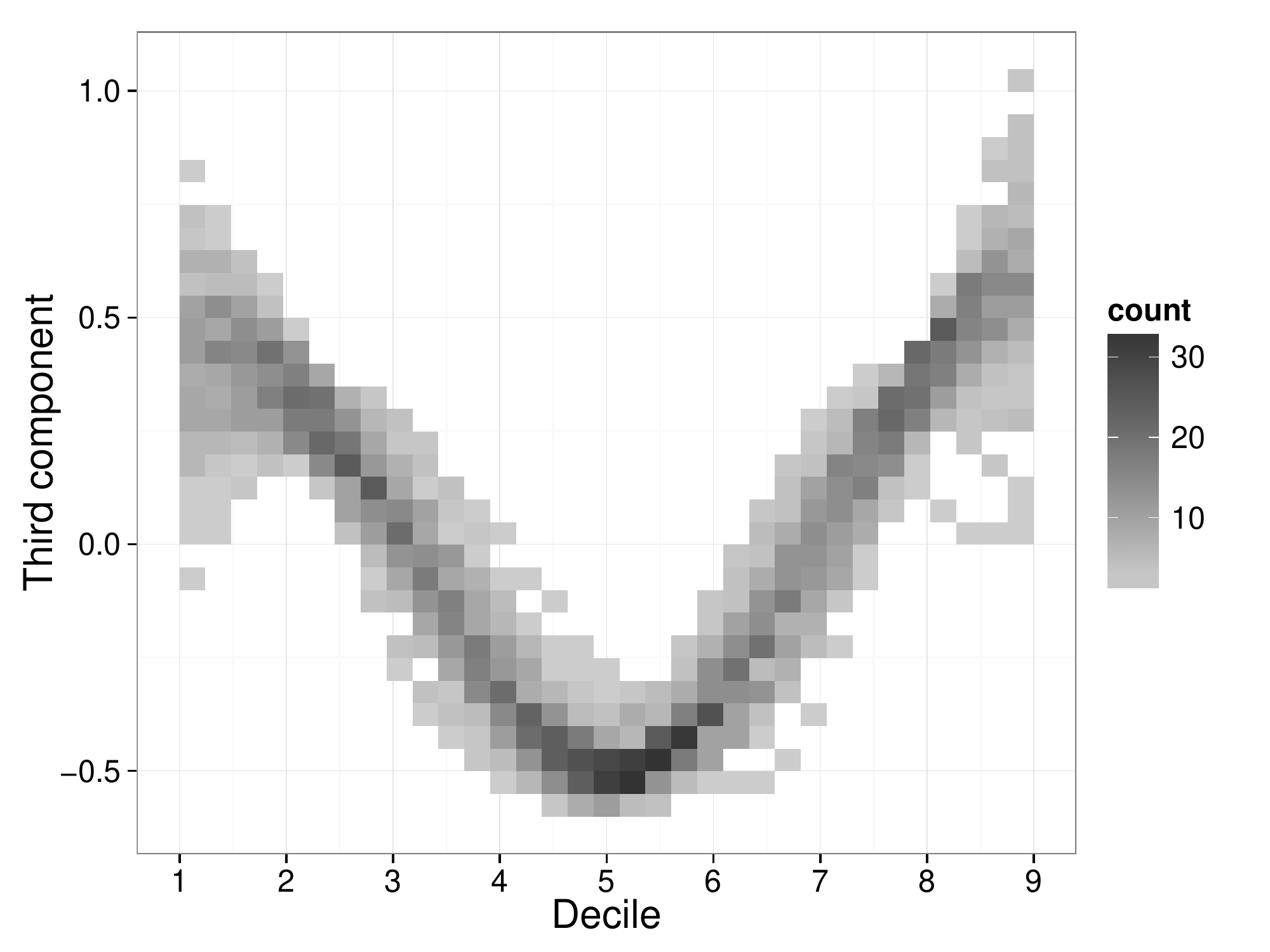}
	\caption{The first three functional PCs extracted from the LRP data every day, projected onto the same axes for the spread (top) and XLM (bottom)}
	\label{fig:pcaharmonics}
	\end{center}
\end{figure}

Specifically, given smoothed functional data $\left\{x_i(u)\right\}_{i \in 1:I}$, we are searching for the weight functions $\xi$, such that the corresponding scores 
\begin{equation}
f_i=\int \xi(u) x_i(u) du
\end{equation}
have the largest possible variation. That is, weight function $\xi_1$ should maximise 
\begin{equation}
\sum_i \left[\int \xi(u) x_i(u) du\right]^2
\end{equation} 
subject to the constraint $\int \xi(u)^2 du=1$. In this context, $\xi_1$ will be the most important functional component of the market-wide liquidity resilience profile and will correspond to the dominant model of functional variation. 

Consider the mean and covariance functions for the functional LRPs denoted by
$\mu(u)=E(x_i(u))$ and $v(u,w)$
\begin{equation}
v(u,w)=\frac{1}{N-1}\sum_{i=1}^{N} x_i(u)x_i(w)
\end{equation}
and the covariance operator V 
\begin{equation}
V\xi=\int v(\cdot,u) \xi(u)du
\end{equation}
This operator has orthonormal eigenfunctions $\xi_1,\ldots,\xi_K$, with eigenvalues $\rho_1 \geq \rho_2 \geq \ldots \geq \rho_K$, satisfying $V\xi_k=\rho_k \xi_k$

In Figure \ref{fig:pcaharmonics} we project the first three leading market liquidity resilience factors from the FPCA from every daily dataset onto the same axes, in order to understand whether the dominant modes of variation vary over time. In the vein of the liquidity commonality literature, we will call these the market factors of resilience. We note that the first FPC is fairly constant over time, and is greater at higher threshold levels for the spread. This indicates that the market component of resilience is important for explaining deviations from more extreme levels of the spread. Once we consider a liquidity measure which takes depth into account, however, such as the XLM, the opposite seems to apply: we observe that the contribution of the first FPC from the daily XLM liquidity resilience profiles tends to decrease at higher thresholds.  

There are two distinct modes for the second FPC, which become almost identical, if one is flipped across the x-axis, and this is the case for both the spread and the XLM. We find that for some assets, multiplying the second PC by the score for that asset almost eliminates the second mode of variation and thus the effect of the second PC becomes relatively constant over time.

\section{Functional principal component regression for LRPs}
\label{sec:fpcaregression}
Recall that \cite{mancini2013liquidity} regress individual exchange rate liquidity against the principal component obtained over all rates, and interpret the $R^2$ coefficient of determination for every asset as the degree of commonality. For our study of liquidity resilience commonality we perform a similar regression idea except we extend this in our case to the functional space setting. The functional principal components obtained every day that characterize the market liquidity resilience factors will now be used as functional covariates, in order to explain the variation in LRPs for individual assets inter-daily.

A linear regression model relating a functional response to a collection of functional covariates at the same points is called a concurrent model and is given as follows:
\begin{equation}
x_{i,t}(u)=\beta_0(u)+\sum_{j=1}^q \beta_j(u) \xi_{j,t}(u) + \epsilon_{i,t}(u)
\end{equation}
where $t$ is the day index, $\beta_0$ is the intercept function, and the $\beta_j$ are coefficient functions of the covariate functions $\xi_j$ i.e. the market functional PCs. $\beta_0$ could also be considered as being the product with a constant function whose value is always one. Let the $t$ by $q$ matrix $\mathbf{Z}$ contain the covariate functions $\xi_{j,t}$. In matrix notation, the concurrent functional linear model is given by
\begin{equation}
\mathbf{x}(u)=\mathbf{Z}(u)\boldsymbol{\beta}(u)+\boldsymbol{\epsilon}(u)
\end{equation} 
and the fitting criterion (if we also include a roughness penalty $J(\beta_j)$) becomes
\begin{equation}
LMSSE(\boldsymbol{\beta})=\int \mathbf{r}(u)'\mathbf{r}(u)du + \sum_{j=1}^q \lambda_j \int J(\beta_j(u))du 
\end{equation}
where $\mathbf{r}(u)=\mathbf{x}(u)-\mathbf{Z}(u)\boldsymbol{\beta}(u)$. For the estimation method utilised see details in \cite{ramsay2006functional}.

\begin{figure}[ht!]
	\begin{center}
	\includegraphics[width=0.4\textwidth]{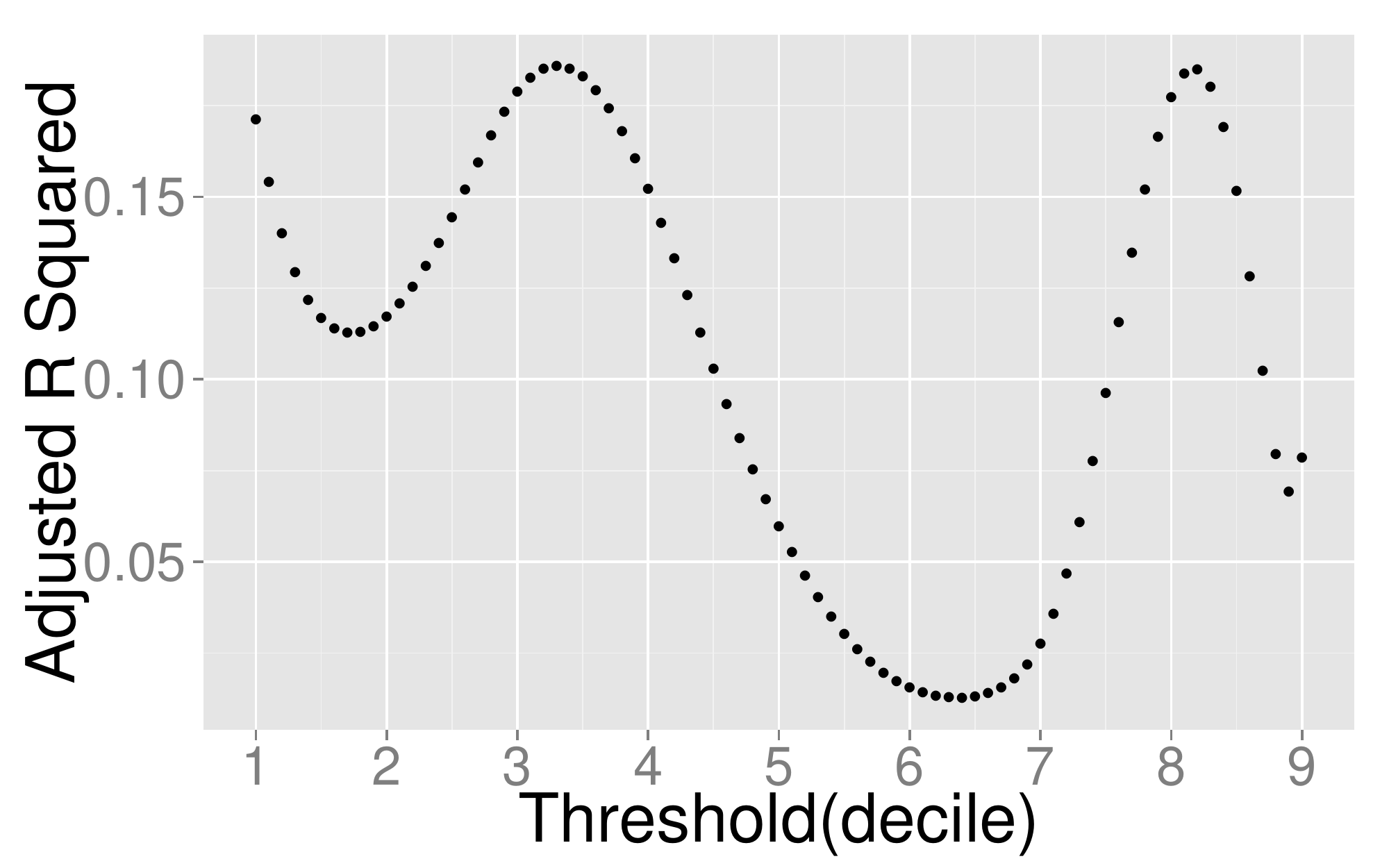}
		\includegraphics[width=0.4\textwidth]{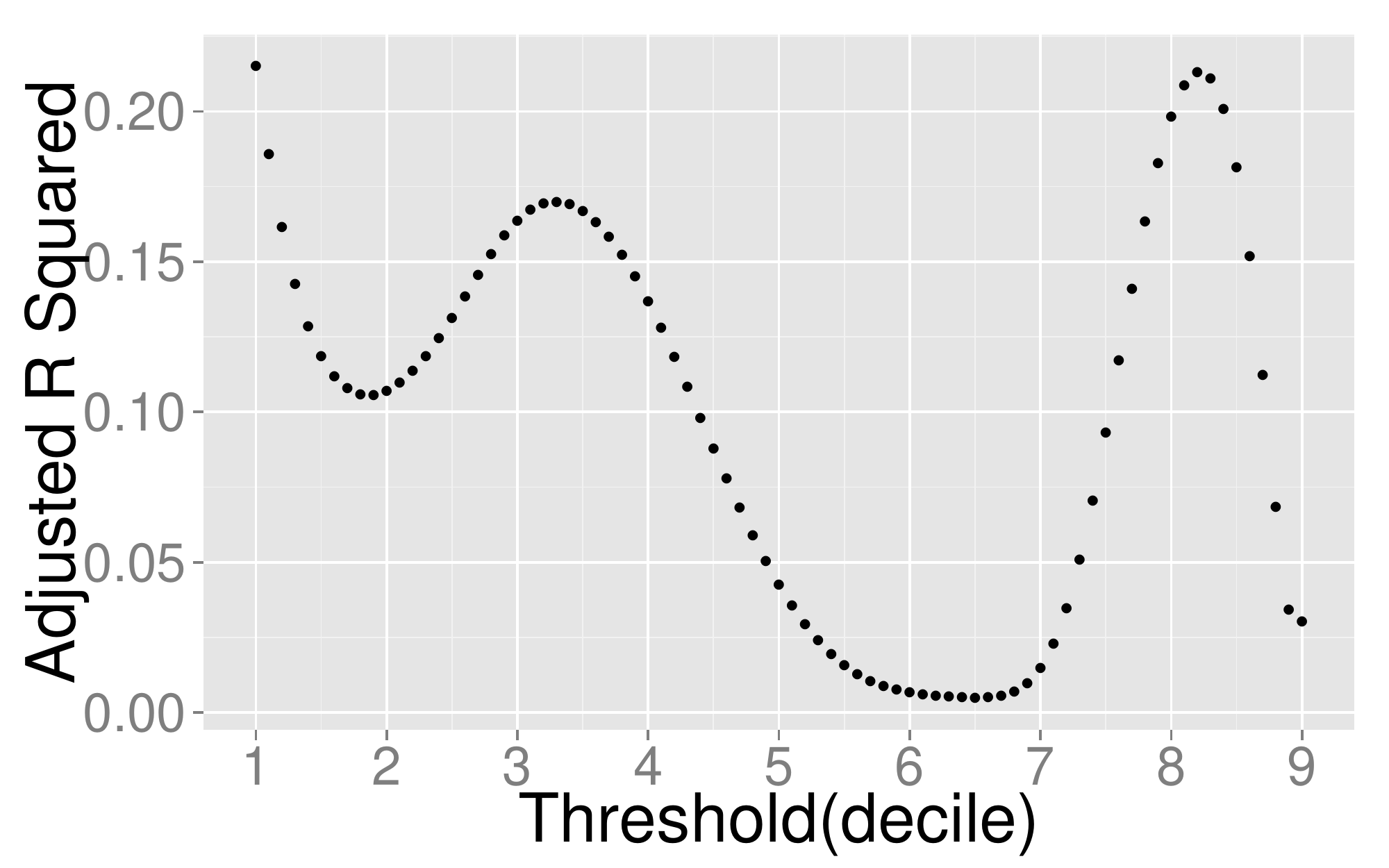}
			\includegraphics[width=0.4\textwidth]{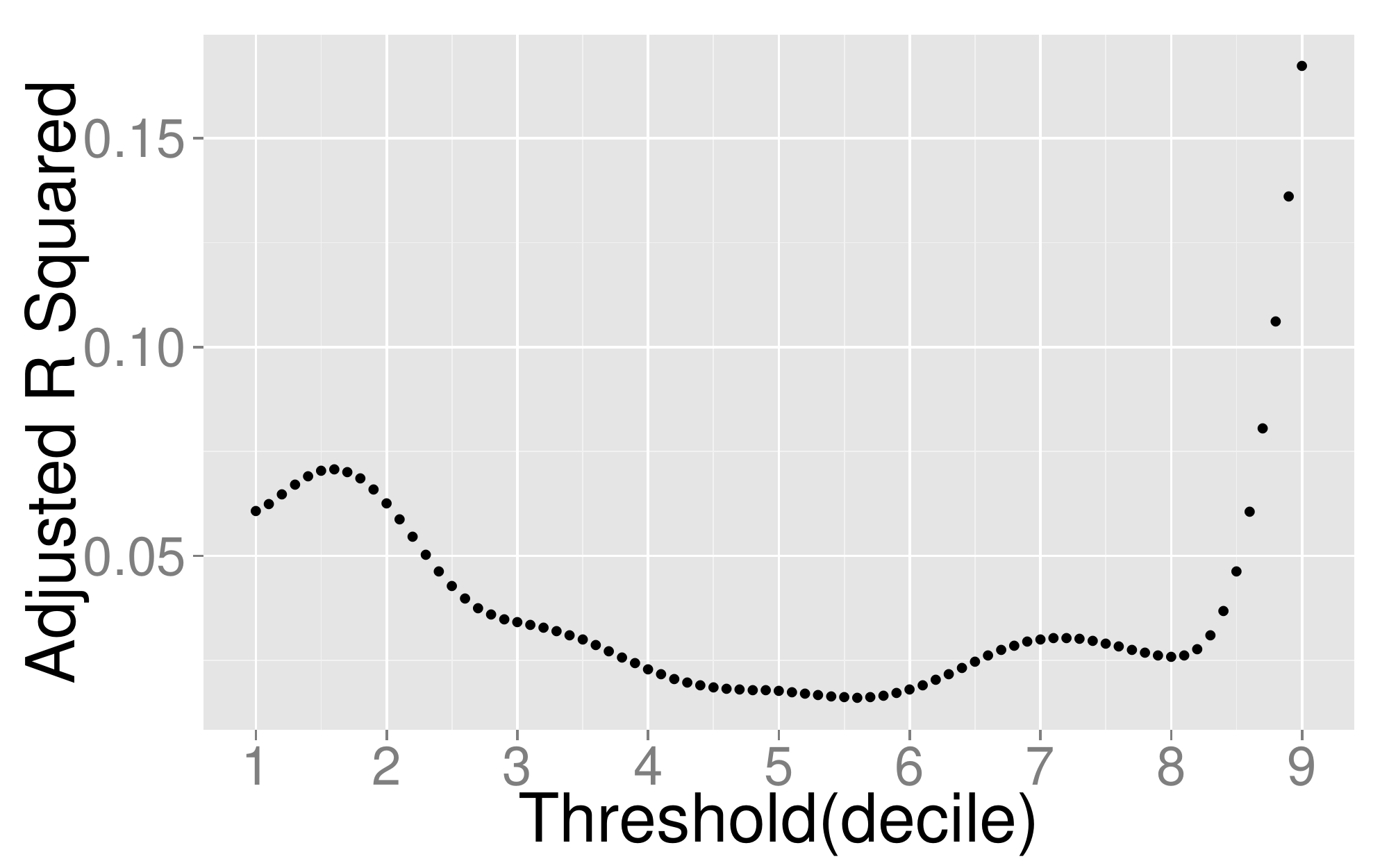}
		\includegraphics[width=0.4\textwidth]{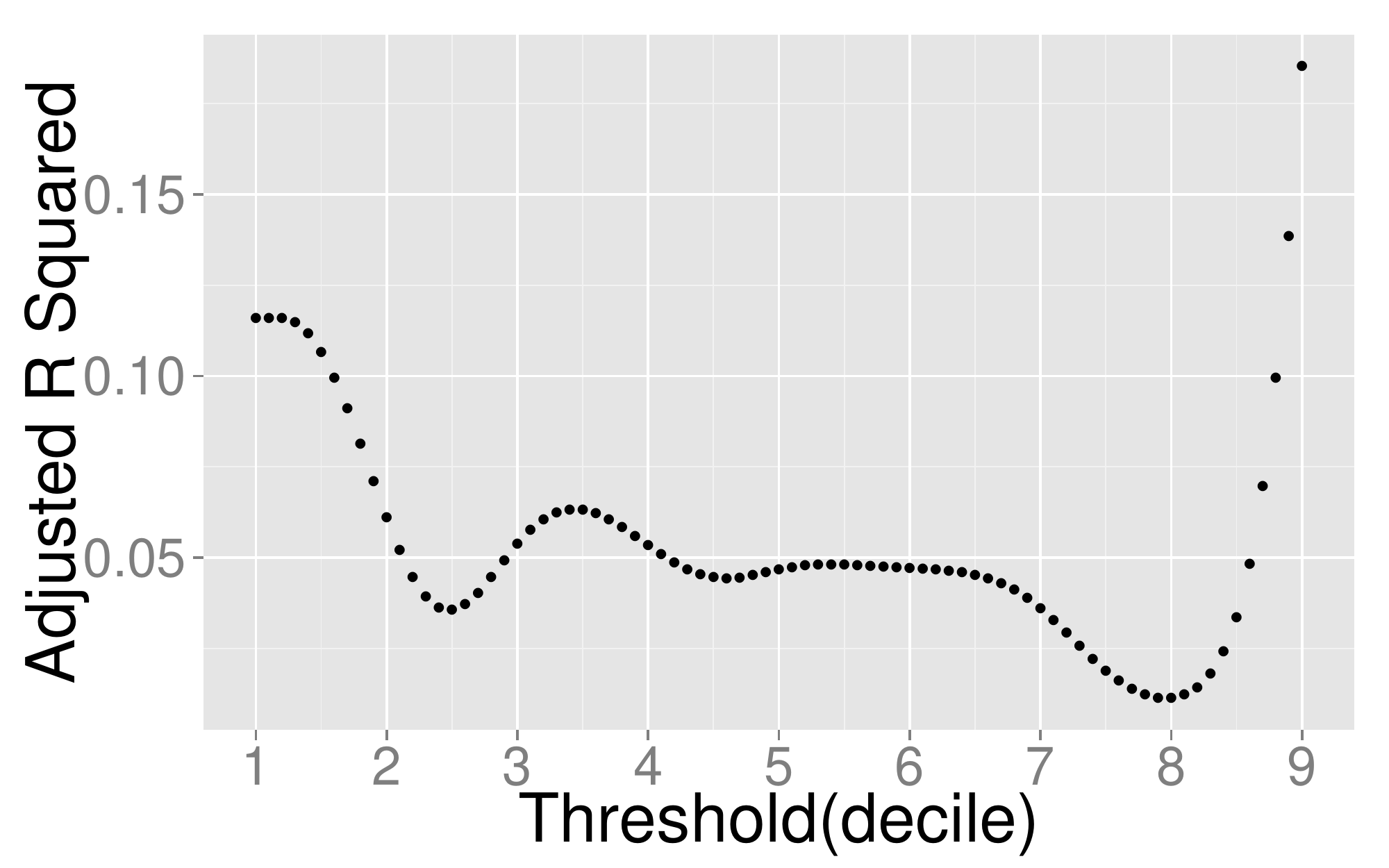}
	\caption{The $R^2$ functions obtained from regressing individual asset Liquidity Resilience Profiles against the first two PCs obtained from the daily LRP curves using the spread (top) and XLM(bottom) for stocks Nexans SA and Credit Agricole.}
	\label{fig:R2}
	\end{center}
\end{figure}

The entire functional PCA regression procedure is summarised in the following steps: 
\begin{enumerate}
\item We first obtain functional representations of the LRP for every asset on every day. 
\item We then extract the first 3 components from the LRPs every day, which will serve as our covariates.
\item We set up a basis for the coefficient functions $\beta_0,\beta_1,\beta_2,\beta_3$. 
\item Finally, we take LRPs for a single asset over time (this will be the dependent variable) and run the regression. 
\end{enumerate}
Here $\beta_0$ will have a constant basis, while for $\beta_1,\beta_2,\beta_3$ we set up a cubic spline basis as before, but with 5 basis functions. We imposed the same $J_2$ roughness penalisation as before, in order to avoid possible overfitting. We can assess the quality of the fit for asset $i$ using the $R^2$ \textit{function}
\begin{equation}
SS_{reg}(u) = \sum_{i,t} [ \hat{x}_{i,t}(u)-\mu_t(u)]^2,SS_{res}(u) = \sum_{i,t} [ \hat{x}_{i,t}(u)-x_{i,t}(u)]^2
\end{equation}

\begin{equation}
R^2(u)=\frac{SS_{reg}(u)}{SS_{reg}(u)+SS_{res}(u)}
\end{equation}

We should note that in our results we omit the intercept function $\beta_0$, as it did not increase the explanatory power of the regression. We present the $R^2$ function for two assets , where we can now observe the explanatory power of the regression at different threshold levels. That is at threshold level $u$, $R^2(u)$ denotes the proportion of variation in the LRP of an individual asset at threshold $u$ that is explained by the first 3 components obtained from the FPCA analysis, again at level $u$. The $R^2$ varies as we alter $u$, and this is consistent across assets, as we can see in Figure \ref{fig:R2}. The advantage of this representation is that we can identify the ranges of $u$ liquidity thresholds for which the principal components are more or less explanatory over time.

\begin{figure}[ht!]
	\begin{center}
	\includegraphics[width=0.49\textwidth]{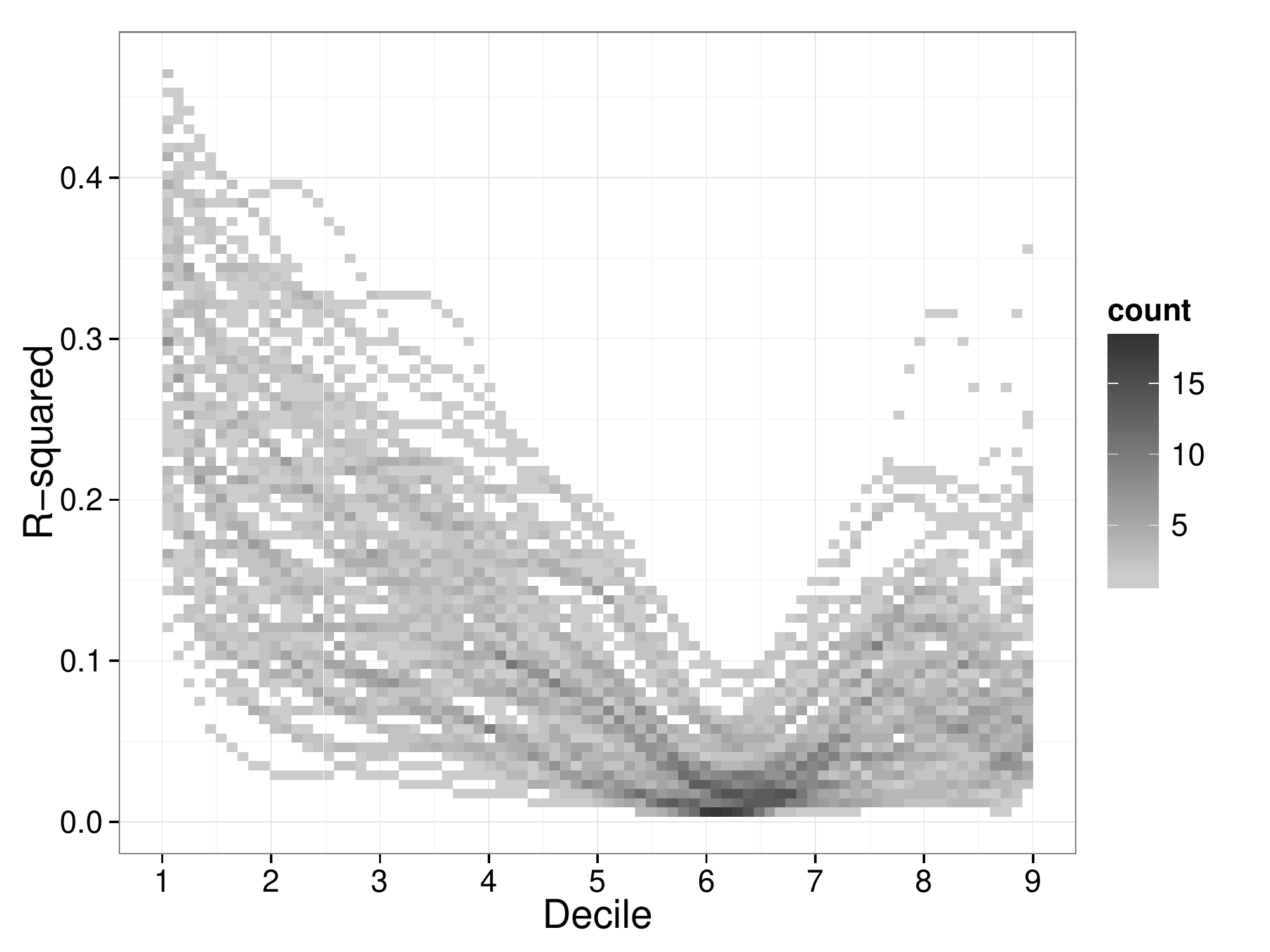}
		\includegraphics[width=0.49\textwidth]{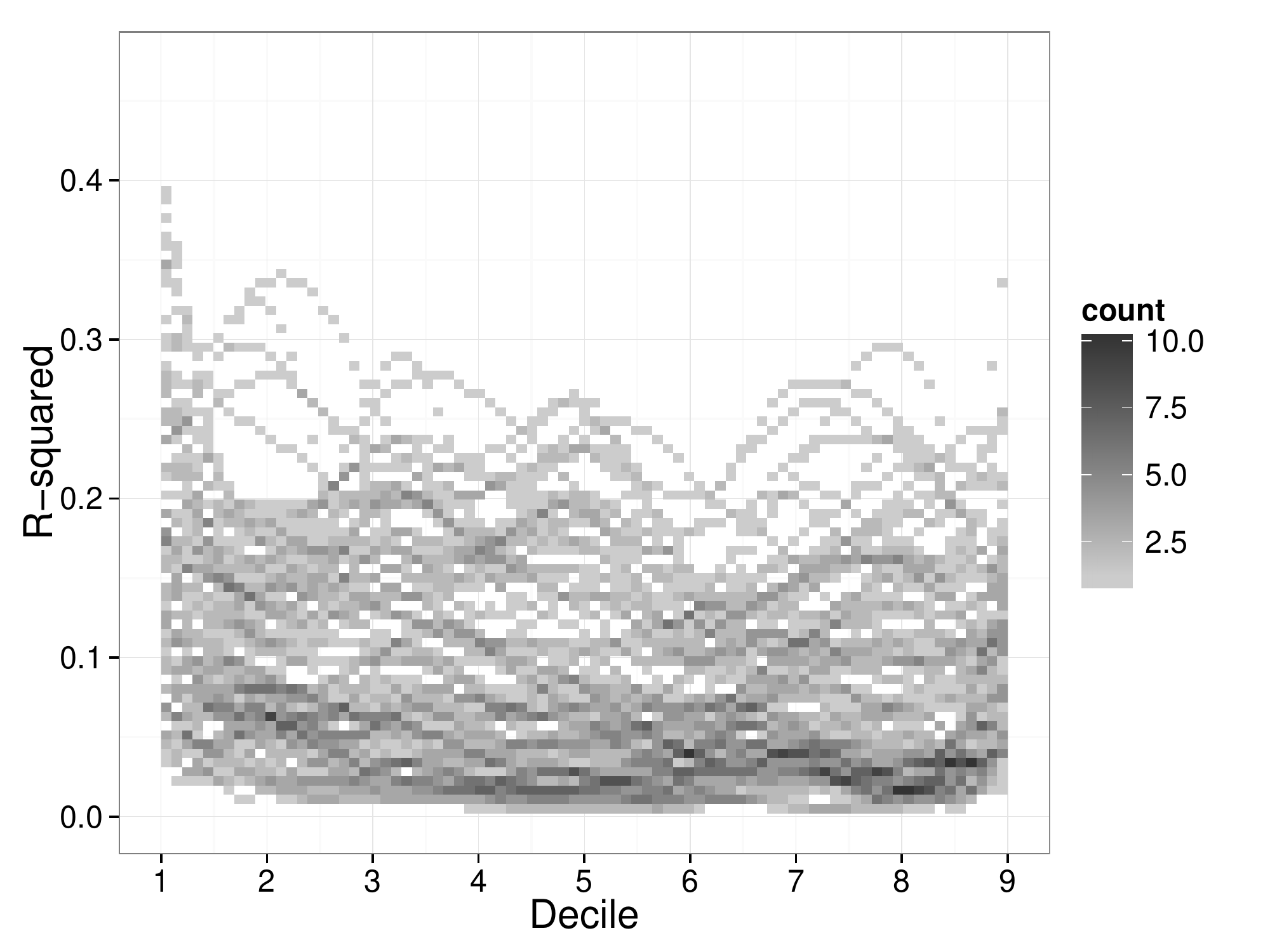}
	\caption{The $R^2$ functions obtained from regressing individual asset Liquidity Resilience Profiles against the first three PCs obtained from the daily LRP curves for the spread(left) and XLM(right).}
	\label{fig:allregr2}
	\end{center}
\end{figure}

We note that in the case of the spread, the PCs are explanatory about the initial part of the curve, that is, where we consider deviations from relatively low spreads. Just as market factors (principal components) extracted from the spreads of individual assets can explain around 25\%of the variation in the absolute level of the spread for these assets (as we had noted in Figure \ref{fig:allr2pca}), PCs extracted from the LRPs can explain between 10 and 40\% of the variation in the expected duration of spread deviations, at least at lower spreads. Once we consider deviations from higher spreads, the explanatory power of market factors drops sharply. 

This indicates that there are additional factors that become important at higher liquidity levels, which are specific for each asset. A possible explanation for this would be a difference in the efficiency of market making between assets. As there are varying requirements for market making for assets in different jurisdictions, it is possible that in certain assets, market makers can stop operating in more illiquid LOB regimes, to avoid building any position that would then be costly to unwind. This would mean that for those assets, the expected duration of deviations above greater levels of the spread would be higher, as there would be fewer market participants willing to replenish the market after a shock.

In the case of the XLM, we find that the commonality in liquidity resilience - which we again measure through the explanatory power of the market factors, extracted through the functional PCA - is markedly lower than the corresponding temporal commonality of the XLM across assets. A possible reason for the lower adjusted $R^2$ levels for the XLM compared to the respective figures for the spread is that while market makers may act similarly in trying to tighten the spread after a shock, they are less inclined to post large volumes in the LOB for certain assets, in order to avoid excess exposure in a market where it will be costly to unwind a position.

\section{Conclusion}
\label{sec:conc}

We have reviewed the performance of the standard approach for measuring liquidity commonality through principal components regression, on a massive dataset from a pan-European equity venue. We have shown that the assumption that one can capture the most important features of liquidity commonality, through methods which are based on second moments, will not always be appropriate. We therefore perform a projection-pursuit based ICA (Independent Component Analysis), which addresses this issue by incorporating higher order information, to assess commonality in liquidity.

The standard approach to liquidity commonality fails to capture commonality in the resilience of liquidity, or the speed of replenishment of the LOB. We addressed this by proposing an approach to quantify the commonality in liquidity resilience (as characterised by \cite{panayi2014market}) by first obtaining a functional representation of resilience and then measuring the explanatory power of market factors extracted from the asset cross-section. We have shown that functional data analysis can be very valuable for characterising features of massive datasets, such as those extracted from high-frequency LOB data, as it can vastly reduce the dimensionality of the data and make comparisons between assets possible.

Our results suggest that market factors for liquidity resilience (as captured by functional principal components analysis) can explain between 10 and 40\% of the time required for the spread to return to a low threshold level after a shock. The same market factors are found to be much less explanatory if we consider higher threshold levels of the spread. Once we also consider a liquidity measure that takes depth into account, such as the XLM, the explanatory power diminishes significantly. 

We have interpreted these results through the prism of market making activity in the LOB. While market makers may act similarly in trying to tighten the spread after a shock, they are less inclined to post large volumes in the LOB for certain assets, in order to avoid excess exposure. We also identified the possible absence of quoting obligations for certain assets to be a contributing factor to explaining these outcomes.

Contrasting these results with our liquidity commonality findings, we find that temporal commonality in the liquidity measures does not necessarily entail commonality in liquidity resilience. We would argue that this has positive implications for market quality, as it indicates that slow liquidity replenishment in certain assets is not necessarily contagious for the market. Future studies will further explore the economic ramifications of these findings in detail.

\bibliographystyle{alpha}
\bibliography{all,rQUFguide}

\newpage
\appendix
\section{Modelling the TED}
\label{sec:TED}

We define the time of exceedance $T_i$ and exceedance duration $\tau_i^{TED}$ formally as follows:

\begin{itemize}
\item $T_i$ will denote the $i$-th random time instant in a trading day that the liquidity measure $LM_{T_{i}}$ upcrosses the threshold $c$, i.e. $T_{i} = \inf\left\{t: LM_{t} > c, \; t > T_{i-1}+\tau_{i-1}^{\text{TED}}, t > T_0  \right\}$.

\item $\tau_i^{\text{TED}}$ is the duration of time in ms that the liquidity measure $LM_{t}$ remains above the threshold $c$, i.e. $\tau_i^{\text{TED}}=\inf\left\{t~:~LM_{T_i+t} \leq c, \; t > 0 \right\}$
\end{itemize}

\cite{panayi2014market} modelled the variation in TED observations using a survival regression model, where they assumed a flexible 3 parameter Generalised Gamma distribution:

\begin{equation*}
\tau_i^\text{TED} \stackrel{i.i.d}{\sim} F\left(t;k,a,b\right) = \frac{\gamma\left(k,\left(\frac{t}{a}\right)^{b}\right)}{\Gamma\left(k\right)}
\end{equation*}

This family of distributions encompasses the Weibull distribution (with $k=1$) and the Lognormal as a limiting case (as $k \rightarrow \infty$). The results we present here are with the Lognormal distribution:

\begin{equation*}
	\log(\tau_i^\text{TED})=\mathbf{x}'_i\boldsymbol{\beta}+\varepsilon_i
\end{equation*}

where $\varepsilon \overset{i.i.d}{\sim} N(0,(\sigma)^2)$ and $\mathbf{x}$ is a vector of covariates from the LOB: 

\begin{itemize}
\item{The total number of asks in the first 5 levels of the LOB at time $t$, obtained according to $x_t^{(1)}=\sum_{i=1}^{5}\left |V_{t}^{a,i}\right |$} (where $\left | \cdot  \right |$ is the number of orders at a particular level)
\item{The total number of bids in the first 5 levels of the LOB at time $t$, obtained according to $x_t^{(2)}=\sum_{i=1}^{5}\left |V_{t}^{b,i}\right |$}
\item{The total ask volume in the first 5 levels of the LOB at time $t$, obtained according to $x_t^{(3)}=\sum_{i=1}^{5}TV_{t}^{a,i}$}
\item{The total bid volume in the first 5 levels of the LOB at time $t$, obtained according to $x_t^{(4)}=\sum_{i=1}^{5}TV_{t}^{b,i}$}
\item The instantaneous value of the liquidity measure (spread or XLM) at the point at which the $i$-th exceedance occurs, which is given by $x_t^{(5)}=P_{t}^{a,1}-P_{t}^{b,1}$
\item The number $x_t^{(6)}$ of previous TED observations in the interval $[t-\delta,t]$,  with $\delta=1s$
\item The time since the last exceedance, $x_t^{(7)}$
\item The average of the last 5 TEDs, $x_t^{(8)}$
\item A dummy variable indicating if the exceedance occurred as a result of a market order to buy, $x_t^{(9)}$
\item A dummy variable indicating if the exceedance occurred as a result of a market order to sell, $x_t^{(10)}$
\end{itemize}

\newpage
\section{Additional figures}
\label{sec:figs}

\begin{figure}[ht!]
	\begin{center}
	\includegraphics[width=0.5\linewidth]{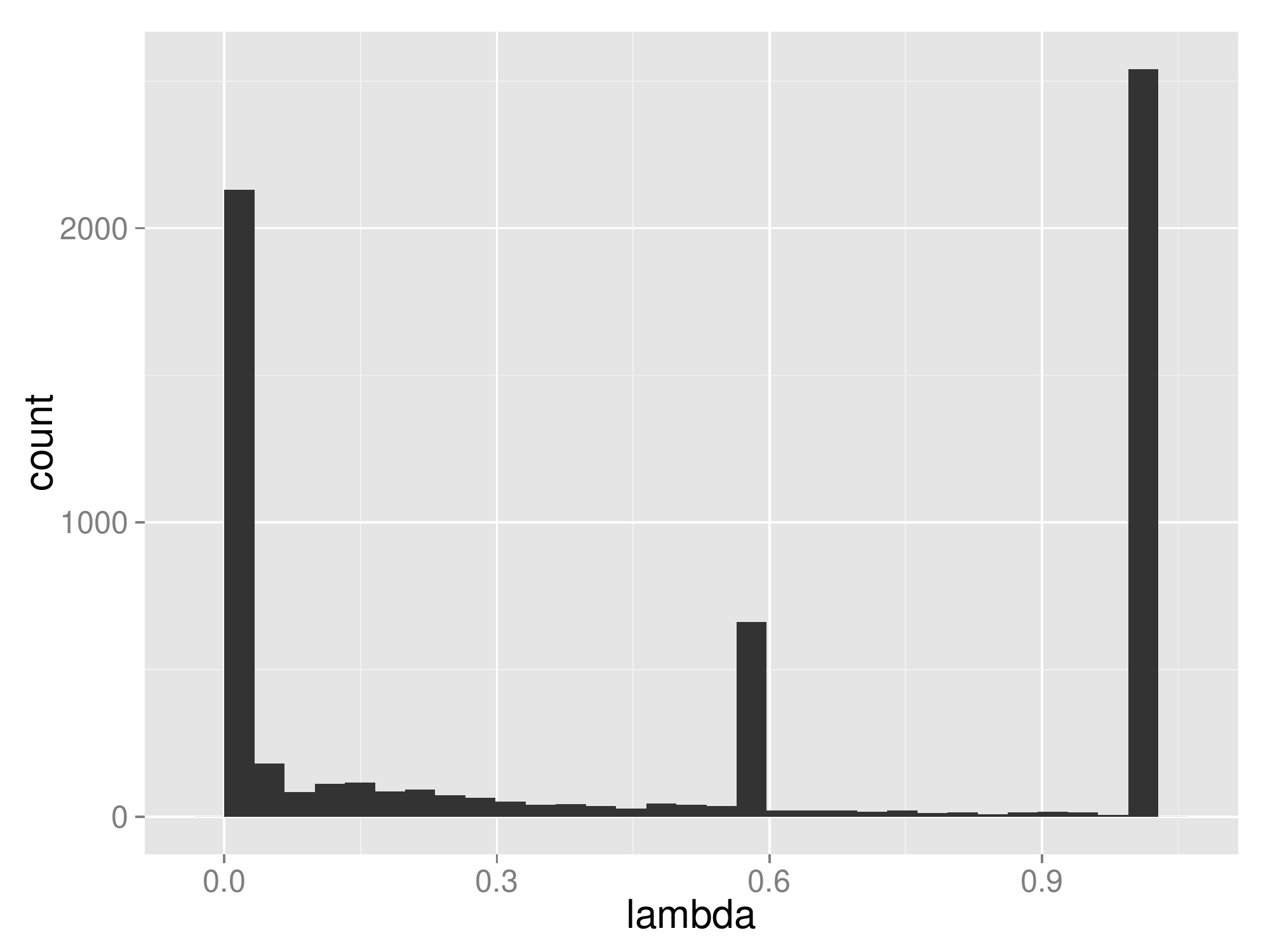}
	\caption{The optimal $\lambda$ value calculated by the GCV procedure for every LRP fit (i.e. for every asset on every day).}
	\label{fig:gcv}
	\end{center}
\end{figure}

\begin{figure}[ht!]
	\begin{center}
	\includegraphics[width=0.49\textwidth]{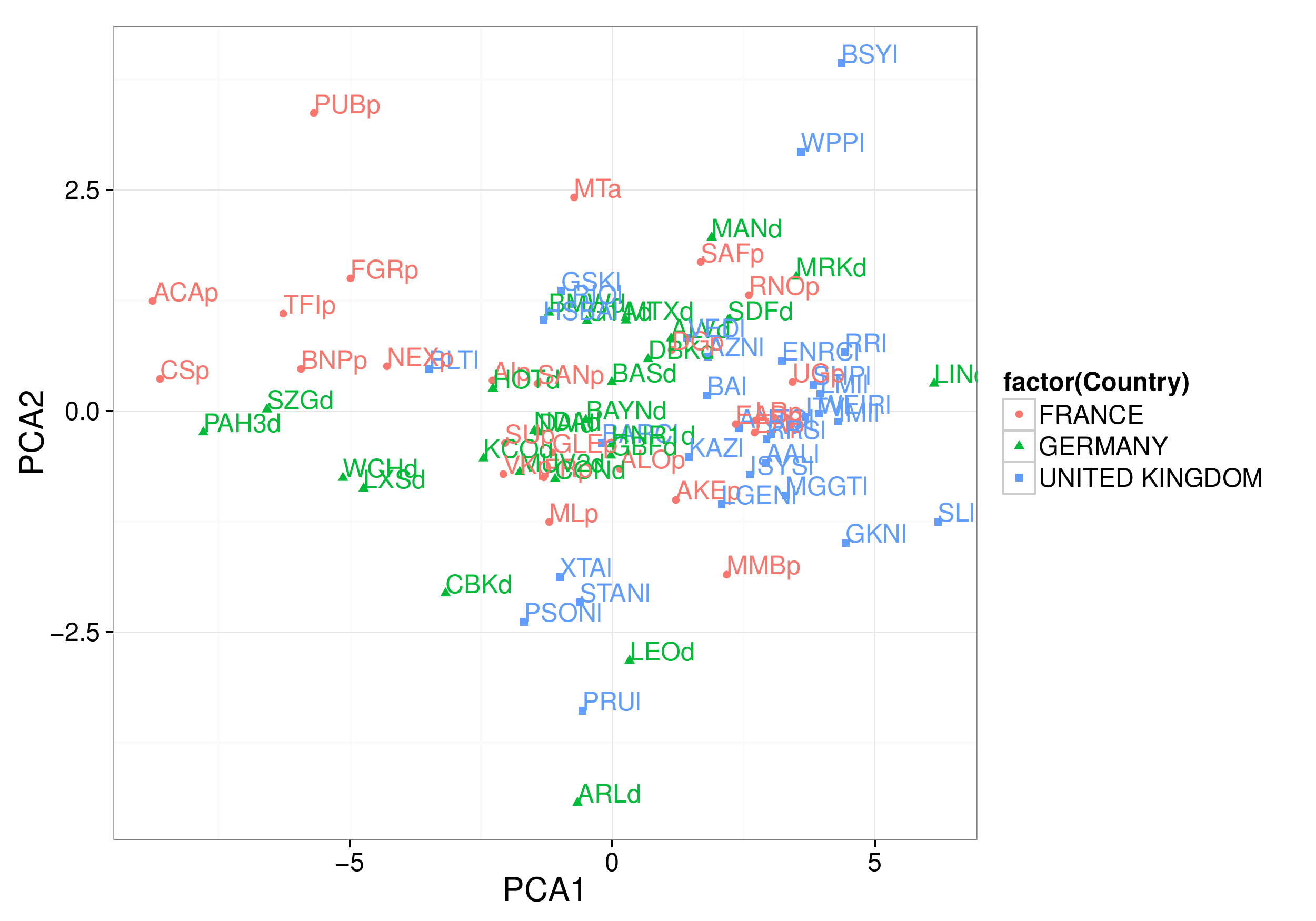}
		\includegraphics[width=0.49\textwidth]{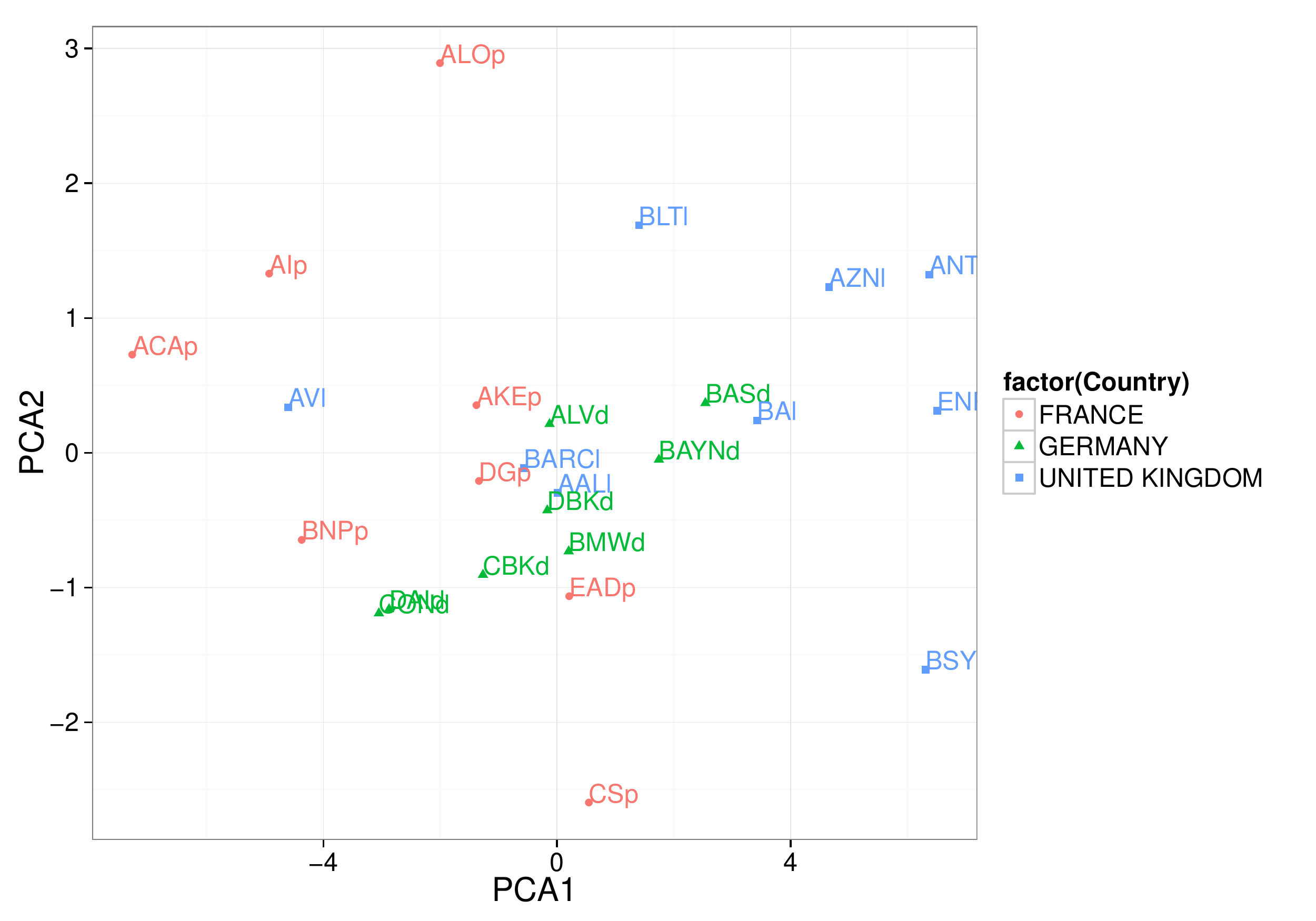}
	\caption{Scores for the first two PCs for every asset for the spread(left) and XLM(right).   
}
	\label{fig:pca}
	\end{center}
\end{figure}

\newpage
\section{Company information}
\label{sec:comp}
\begin{table}[ht]
\centering
\resizebox{0.6\linewidth}{!}{
\begin{tabular}{rllll}
  \hline
 & Country & Name & Symbol & Sector \\ 
  \hline
1 & FRANCE & EADS (PAR) & EADp & Aerospace / Defense \\ 
  2 & FRANCE & SAFRAN & SAFp & Aerospace / Defense \\ 
  3 & FRANCE & VALEO & FRp & Automobiles / Auto Parts \\ 
  4 & FRANCE & MICHELIN & MLp & Automobiles / Auto Parts \\ 
  5 & FRANCE & RENAULT & RNOp & Automobiles / Auto Parts \\ 
  6 & FRANCE & PEUGEOT & UGp & Automobiles / Auto Parts \\ 
  7 & FRANCE & CREDIT AGRICOLE & ACAp & Banking Services \\ 
  8 & FRANCE & BNP PARIBAS & BNPp & Banking Services \\ 
  9 & FRANCE & SOCIETE GENERALE & GLEp & Banking Services \\ 
  10 & FRANCE & SANOFI & SANp & Biotechnology / Pharmaceuticals \\ 
  11 & FRANCE & AIR LIQUIDE & AIp & Chemicals \\ 
  12 & FRANCE & ARKEMA & AKEp & Chemicals \\ 
  13 & FRANCE & VINCI (EX SGE) & DGp & Construction / Engineering / Materials als \\ 
  14 & FRANCE & BOUYGUES & ENp & Construction / Engineering / Materials als \\ 
  15 & FRANCE & EIFFAGE & FGRp & Construction / Engineering / Materials als \\ 
  16 & FRANCE & AXA & CSp & Insurance \\ 
  17 & FRANCE & ALSTOM & ALOp & Machinery / Equipment / Components \\ 
  18 & FRANCE & LEGRAND & LRp & Machinery / Equipment / Components \\ 
  19 & FRANCE & NEXANS & NEXp & Machinery / Equipment / Components \\ 
  20 & FRANCE & SCHNEIDER ELECTRIC & SUp & Machinery / Equipment / Components \\ 
  21 & FRANCE & LAGARDERE GROUPE & MMBp & Media / Publishing \\ 
  22 & FRANCE & PUBLICIS GROUPE & PUBp & Media / Publishing \\ 
  23 & FRANCE & TF1 (TV.FSE.1) & TFIp & Media / Publishing \\ 
  24 & FRANCE & ARCELORMITTAL & MTa & Metal / Mining \\ 
  25 & FRANCE & VALLOUREC & VKp & Metal / Mining \\ 
  26 & GERMANY & MTU AERO ENGINES HLDG. & MTXd & Aerospace / Defense \\ 
  27 & GERMANY & BMW & BMWd & Automobiles / Auto Parts \\ 
  28 & GERMANY & CONTINENTAL & CONd & Automobiles / Auto Parts \\ 
  29 & GERMANY & DAIMLER & DAId & Automobiles / Auto Parts \\ 
  30 & GERMANY & PORSCHE AML.HLDG.PREF. & PAH3d & Automobiles / Auto Parts \\ 
  31 & GERMANY & AAREAL BANK & ARLd & Banking Services \\ 
  32 & GERMANY & COMMERZBANK & CBKd & Banking Services \\ 
  33 & GERMANY & DEUTSCHE BANK & DBKd & Banking Services \\ 
  34 & GERMANY & BAYER & BAYNd & Biotechnology / Pharmaceuticals \\ 
  35 & GERMANY & MERCK KGAA & MRKd & Biotechnology / Pharmaceuticals \\ 
  36 & GERMANY & BASF & BASd & Chemicals \\ 
  37 & GERMANY & LINDE & LINd & Chemicals \\ 
  38 & GERMANY & LANXESS & LXSd & Chemicals \\ 
  39 & GERMANY & K + S & SDFd & Chemicals \\ 
  40 & GERMANY & WACKER CHEMIE & WCHd & Chemicals \\ 
  41 & GERMANY & GEA GROUP & G1Ad & Construction / Engineering / Materials als \\ 
  42 & GERMANY & BILFINGER BERGER & GBFd & Construction / Engineering / Materials als \\ 
  43 & GERMANY & HOCHTIEF & HOTd & Construction / Engineering / Materials als \\ 
  44 & GERMANY & ALLIANZ & ALVd & Insurance \\ 
  45 & GERMANY & HANNOVER RUCK. & HNR1d & Insurance \\ 
  46 & GERMANY & MUENCHENER RUCK. & MUV2d & Insurance \\ 
  47 & GERMANY & LEONI & LEOd & Machinery / Equipment / Components \\ 
  48 & GERMANY & MAN & MANd & Machinery / Equipment / Components \\ 
  49 & GERMANY & KLOECKNER \& CO & KCOd & Metal / Mining \\ 
  50 & GERMANY & AURUBIS & NDAd & Metal / Mining \\ 
  51 & GERMANY & SALZGITTER & SZGd & Metal / Mining \\ 
  52 & UK & BAE SYSTEMS & BAl & Aerospace / Defense \\ 
  53 & UK & MEGGITT & MGGTl & Aerospace / Defense \\ 
  54 & UK & ROLLS-ROYCE HOLDINGS & RRl & Aerospace / Defense \\ 
  55 & UK & GKN & GKNl & Automobiles / Auto Parts \\ 
  56 & UK & BARCLAYS & BARCl & Banking Services \\ 
  57 & UK & HSBC HDG. & HSBAl & Banking Services \\ 
  58 & UK & STANDARD CHARTERED & STANl & Banking Services \\ 
  59 & UK & ASTRAZENECA & AZNl & Biotechnology / Pharmaceuticals \\ 
  60 & UK & GLAXOSMITHKLINE & GSKl & Biotechnology / Pharmaceuticals \\ 
  61 & UK & SHIRE & SHPl & Biotechnology / Pharmaceuticals \\ 
  62 & UK & AVIVA & AVl & Insurance \\ 
  63 & UK & LEGAL \& GENERAL & LGENl & Insurance \\ 
  64 & UK & PRUDENTIAL & PRUl & Insurance \\ 
  65 & UK & STANDARD LIFE & SLl & Insurance \\ 
  66 & UK & IMI & IMIl & Machinery / Equipment / Components \\ 
  67 & UK & INVENSYS & ISYSl & Machinery / Equipment / Components \\ 
  68 & UK & WEIR GROUP & WEIRl & Machinery / Equipment / Components \\ 
  69 & UK & BRITISH SKY BCAST.GROUP & BSYl & Media / Publishing \\ 
  70 & UK & ITV & ITVl & Media / Publishing \\ 
  71 & UK & PEARSON & PSONl & Media / Publishing \\ 
  72 & UK & WPP & WPPl & Media / Publishing \\ 
  73 & UK & ANGLO AMERICAN & AALl & Metal / Mining \\ 
  74 & UK & ANTOFAGASTA & ANTOl & Metal / Mining \\ 
  75 & UK & BHP BILLITON & BLTl & Metal / Mining \\ 
  76 & UK & EURASIAN NATRES.CORP. & ENRCl & Metal / Mining \\ 
  77 & UK & KAZAKHMYS & KAZl & Metal / Mining \\ 
  78 & UK & LONMIN & LMIl & Metal / Mining \\ 
  79 & UK & RIO TINTO & RIOl & Metal / Mining \\ 
  80 & UK & RANDGOLD RESOURCES & RRSl & Metal / Mining \\ 
  81 & UK & VEDANTA RESOURCES & VEDl & Metal / Mining \\ 
  82 & UK & XSTRATA & XTAl & Metal / Mining \\ 
   \hline
\end{tabular}}
\caption{Country and sector information about the 82 assets used.}
\label{tab:assets}
\end{table}

\end{document}